\documentclass[twocolumn,twocolappendix]{aastex631} 
\usepackage{afterpage}
\usepackage{placeins}
\usepackage{amsmath}
\usepackage{stmaryrd}
\usepackage{booktabs}
\usepackage{multirow}
\usepackage{comment}

\usepackage{soul}

\usepackage{xspace}
\newcommand{\vide}{\tt VIDE\normalfont\xspace}

\begin{document}

\title{Cosmology with voids from the Nancy Grace Roman Space Telescope}

\correspondingauthor{Giovanni Verza}
\email{gverza@flatironinstitute.org}

\author[0000-0002-1886-8348]{Giovanni Verza}
\affiliation{Center for Computational Astrophysics, Flatiron Institute, 162 5$^{\rm th}$ Avenue, 10010, New York, NY, USA}
\affiliation{Center for Cosmology and Particle Physics, Department of Physics, New York University, 726 Broadway, New York, NY 10003, USA}

\author[0009-0001-4912-1087]{Giulia Degni}
\affiliation{Dipartimento di Fisica, Università di Roma Tre, Via della Vasca Navale 84, I-00146 Roma,
Italy}
\affiliation{INFN - Sezione di Roma Tre, Via della Vasca Navale 84, I-00146 Roma, Italy}

\author[0000-0002-6146-4437]{Alice Pisani}

\affiliation{Aix-Marseille Universit\'e, CNRS/IN2P3, CPPM, Marseille, France}
\affiliation{Department of Astrophysical Sciences, Peyton Hall, Princeton University, Princeton, NJ 08544, USA}

\author[0000-0002-0876-2101]{Nico Hamaus}
\affiliation{Universit\"ats-Sternwarte M\"unchen, Fakult\"at f\"ur Physik, Ludwig-Maximilians Universit\"at, Scheinerstr. 1, 81679 M\"unchen, Germany}

\author[0000-0002-0637-8042]{Elena Massara}
\affiliation{Waterloo Centre for Astrophysics, University of Waterloo, 200 University Ave W, Waterloo, ON N2L 3G1, Canada}
\affiliation{Department of Physics and Astronomy, University of Waterloo, 200 University Ave W, Waterloo, ON N2L 3G1, Canada}

\author[0000-0001-5501-6008]{Andrew Benson}
\affiliation{Carnegie Science, 813 Santa Barbara St., Pasadena, CA, USA}

\author[0000-0002-2847-7498]{Stéphanie Escoffier}
\affiliation{Aix-Marseille Universit\'e, CNRS/IN2P3, CPPM, Marseille, France}

\author[0000-0002-4749-2984]{Yun Wang}
\affiliation{IPAC, California Institute of Technology, Mail Code 314-6, 1200 E. California Blvd., Pasadena, CA 91125}

\author[0000-0001-7984-5476]{Zhongxu Zhai}
\affiliation{Department of Astronomy, School of Physics and Astronomy, Shanghai Jiao Tong University, Shanghai 200240, China}
\affiliation{Shanghai Key Laboratory for Particle Physics and Cosmology, Shanghai 200240, China}
\affiliation{Waterloo Centre for Astrophysics, University of Waterloo, 200 University Ave W, Waterloo, ON N2L 3G1, Canada}
\affiliation{Department of Physics and Astronomy, University of Waterloo, 200 University Ave W, Waterloo, ON N2L 3G1, Canada}

\author[0000-0001-7432-2932]{Olivier Doré}
\affiliation{Jet Propulsion Laboratory, California Institute of Technology, Pasadena, CA 91109, USA}
\affiliation{Cahill Center for Astronomy and Astrophysics, California Institute of Technology, Pasadena, CA 91125, USA}

\begin{abstract}
We provide an accurate forecast of the expected constraining power from the main void statistics---the void size function and the void-galaxy cross-correlation function---to be measured by the Roman reference High Latitude Spectroscopic Survey from the Nancy Grace Roman Space Telescope. Relying on a realistic galaxy mock lightcone, covering 2000 square degrees, we find more than $8\times 10^4 $ voids and explore their constraining power in the framework of three different cosmological models: $\Lambda$CDM, $w$CDM, and $w_0 w_{\rm a}$CDM. 
This work confirms the strong complementarity of different void statistics and showcases the constraining power to be expected from Roman voids thanks to the combination of its high tracer density and large observed volume.

\end{abstract}

\section{Introduction} \label{sec:intro}
Cosmic voids, the under-dense regions in the galaxy distribution, provide tight constraints on cosmological parameters~\citep{hamaus_2016,hamaus_2017,sahlen_2016,mao_2017,hawken_2017,achitouv_2017,achitouv_2019,sahlen_2019,hawken_2020,hamaus_2020,nadathur_percival_2020,nadathur_2020,aubert_2022,woodfinden_2022,woodfinden_2023,contarini_2023,fraser_2024,contarini_2024}. Thanks to their wide range of sizes (from tens to hundreds of Mpc), voids allow to access information beyond the two-point correlation function over different scales~\citep{bayer_2021,kreisch_2022,pellicciari_2023,beyond2pt} and promise to provide tight constraints from large scale modern surveys~\citep[e.g.,][]{vsf_euclid,hamaus_2022,radinovic_2023,bonici_2023}. Void statistics greatly benefit of a large observed cosmological volume paired with a small tracer separation, since they allow to extract information down to the smallest scales. 

The Roman reference High Latitude Spectroscopic Survey (Roman reference HLSS) from the Nancy Grace Roman Space Telescope, expected to launch no later than May 2027, will cover in its reference design mission 2,000 square degrees with an unprecedented high tracer number density over such a volume\footnote{We note that, while the Roman reference HLSS covers 2000 square degrees \citep{spergel_2015,wang_2022}, the actual survey that the Nancy Grace Roman Space telescope will execute will be decided in an open community process, and may cover an even greater cosmic volume.}, therefore providing for the first time a cosmic void sample of exceptional quality down to a few Mpcs. Due to the fact that void sizes span a wide range of scales and that a high tracer number density allows a deep sampling of voids, the target population of Roman voids will be complementary to the void populations probed by other past and ongoing spectroscopic galaxy large-scale surveys \citep[see e.g.,][]{hamaus_2016,hamaus_2020,aubert_2022,woodfinden_2022,contarini_2023,thiele_2024}, effectively opening a new window for cosmic void science.

Voids have already shown their power to constrain e.g., $\Omega_m$ and the growth rate of structures $f$ (combined with the galaxy bias or $\sigma_8$)~\citep[e.g.,][]{hamaus_2020}, or the sum of neutrino masses \citep{thiele_2024} with only a few thousand voids from the BOSS sample. The Roman voids upcoming dataset will unlock further constraints, in particular it has been shown that voids can place powerful constraints on the properties of dark energy~\citep{pisani_2015,verza_2019,verza_2023}, and neutrinos~\citep{massara_2015,kreisch_2019,schuster_2019,bayer_2021,kreisch_2022,verza_2023}. Their sensitivity to dark energy properties is expected, since voids are the first regions to be dominated by dark energy~\citep{lavaux_2012,bos_2012,pisani_2015}. Voids are also particularly sensitive to the sum of neutrino masses, since the relative neutrino density in voids is higher than in other environments \citep{schuster_2019,bayer_2024}. Also, neutrinos’ free-streaming scales for relevant values of the sum of neutrino masses correspond to the range of typical void sizes \citep{kreisch_2019}.

When observing voids from large-scale surveys, we can build various statistics. The most mature void statistic is the void-galaxy cross-correlation function (VGCF), that can be used to perform the \citet[AP,][]{AP_1979} test and to measure redshift-space distortions around voids. This statistic currently provides the tighest constraints from voids. Secondly, in recent years the void size function (VSF), providing the number density of voids as a function of their size, has shown sensitivity to a range of cosmological parameters~\citep{pisani_2015,verza_2019,verza_2023, vsf_euclid,contarini_2023,contarini_2024}, as well as a powerful complementarity with other probes~\citep[such as clusters, see][]{bayer_2021,kreisch_2022,pellicciari_2023}. 
Relying on the realistic H$\alpha$ mock from the 2000 square degrees galaxy lightcone~\citep{zhai_2021}, this paper provides a detailed forecast of constraints to be obtained from voids measured thanks to the Nancy Grace Roman Space Telescope.

The paper is organized as follows. In Section~\ref{sec:simu_vide} we introduce the simulation and the void finder, and in Section~\ref{sec:vsf} we present the VSF analysis. Section~\ref{sec:vgcf} describes the VGCF analysis, and we finally conclude in Section~\ref{sec:conclu} summarizing our results and discussing future developments.

\section{Simulations and void finder}\label{sec:simu_vide}

In this work, we consider voids measured in the 2000 square degrees galaxy lightcone~\citep{zhai_2021}, in the redshift range $1<z<2$. The lightcone simulates the H$\alpha$ galaxy redshift catalog expected from the Roman reference HLSS, and is constructed from the {\tt unit} simulation~\citep{chuang_2019}, characterized by a flat $\Lambda$CDM cosmology with parameters: $[h, \Omega_{\rm b}, \Omega_{\rm m}, \sigma_8,n_{\rm s}, A_{\rm s}] = [0.6774, 0.0462, 0.3089, 0.8147, 0.9667, 2.06~\times~10^{-9}$], as measured by~\citet{Planck_2015_params}. The lightcone has been populated with galaxies using the {\tt GALACTICUS}~\citep{benson_2012} semi-analytical model (SAM). The emission line luminosities are evaluated with {\tt CLOUDY}~\citep{ferland_2013,merson_2018} using the ionizing student spectrum of each galaxy as predicted by the SAM as input.

We identified voids in the galaxy lightcone described in ~\cite{zhai_2021} with the \href{https://bitbucket.org/cosmicvoids/vide_public/src/master/}{\vide} public toolkit\footnote{ \url{https://bitbucket.org/cosmicvoids/vide\_public/wiki/Home}}~\citep{sutter_2015_vide}. \vide is a parameter-free topological void finder, which identifies minima and their surrounding basins using the Voronoi tessellation followed by the watershed transform~\citep{schaap_2000_dtfe,platen_2007}, based on {\tt ZOBOV}~\citep{neyrinck_2008_zobov}. The final \vide void catalog provides many void features, the most relevant for our work being i) the volume weighted barycenter, that is the void barycenter obtained weighting by the volumes of the contributing Voronoi cells; ii) the effective radius, $R_{\rm eff}$, i.e. the radius of a sphere with the same volume as the void, $R_{\rm eff} = \left[ (3/4 \pi)  \sum_i V_i \right]^{1/3}$, where $V_i$ is the volume of the $i^{\rm th}$ Voronoi cell belonging to the void.

\begin{figure}[t]
\centering
\includegraphics[width=\columnwidth]{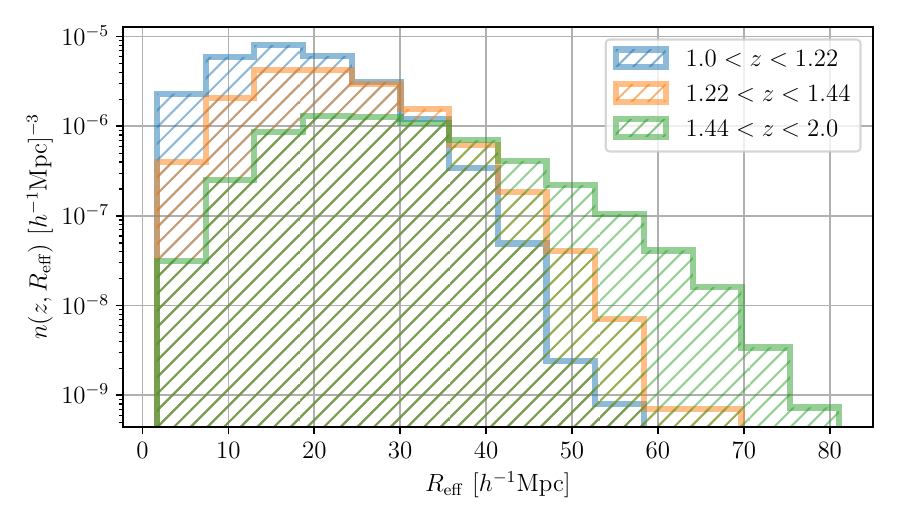}
\caption{Histograms showing the number density of \vide voids as a function of the void effective radius, $R_{\rm eff}$. Each color corresponds to a different redshift bin, as listed in the legend.}
\label{fig:vide_histo}
\end{figure}

We detects 82551 voids in the 2000 squared degrees galaxy lightcone. The redshift binning is chosen to have equipopulated bins for the VGCF analysis, that directly uses the \vide catalog, pruning small voids (see Section~\ref{sec:vgcf} for details). Histograms in Figure~\ref{fig:vide_histo} show the number density of \vide voids as a function of the void effective radius, $R_{\rm eff}$ in each redshift bin considered in the analysis. The first redshift bin, $z \in [1,1.22)$, contains 33574 voids (blue histogram), the second bin, $z \in [1.22,1.44)$, 23114 voids (orange histogram), and the third bin, $z \in [1.44,2]$, 25863 voids (green histogram). 
The higher the number density of galaxies is, the smaller the size of voids that void finders can detect over such a biased tracer distribution~\citep{verza_2023}. 
This is visible in Figure~\ref{fig:vide_histo} as the evolution of the radius distribution as a function of redshift.

\section{The void size function}\label{sec:vsf}

\subsection{Catalog preparation}\label{subsec:vsf_cat}

The theoretical VSF describes the distribution of voids reaching a threshold value in their density contrast field. 
Therefore, in order to compare data with an analytical VSF model, voids are post-processed to have a fixed value in their mean density contrast in the tracer distribution where they are detected~\citep{contarini_2019,verza_2019,vsf_euclid,contarini_2023,contarini_2024,beyond2pt}. 

In this work, we explore a new methodology to post-process the void catalog. From now on, we refer to the post-processed voids as threshold voids, to be distinguished from the watershed voids of the \vide original catalog. The standard methodology for post-processing watershed voids for VSF analyses~\citep{contarini_2019,verza_2019,vsf_euclid,contarini_2023,contarini_2024} consists of growing a sphere around each void center, finding the maximum radius at which the mean density reaches the void formation threshold in the galaxy distribution\footnote{This methodology is conceptually similar to the spherical overdensity package of the {\tt Subfind} algorithm used to post-process Friend-of-Friends halos~\citep{springel_2001_gadget,dolag_2009}}. This methodology represents a simple and robust way to post-process \vide voids, however it presents some arbitrariness.
The first one regards the choice of the void center used to grow the sphere. Usually, the standard choice is to adopt the volume-weighted barycenter provided by \vide. This quantity is the geometrical center of the entire watershed void, and it has been shown to be ideal for measuring the VGCF~\citep{hamaus_2014_profile,hamaus_2016,hamaus_2020,hamaus_2022}. However, it is not guaranteed to be as optimal for threshold voids, which consider only the inner part of voids. The other potential issue concerns the fact that the method grows spheres, and this may introduce dependencies on the cosmology adopted to produce the catalog, a topic that recent analyses are investigating~\citep[][see Section~\ref{subsec:vsf_analysis}]{correa_2021,radinovic_2024}. Moreover, overlaps among the volume of different spherical voids usually occur, leading to the introduction of criteria to select overlapping voids~\citep[usually treated as hard spheres,][]{contarini_2019,verza_2019,vsf_euclid}.
Finally, a spherical void finder step removes the original void shape information coming from the Voronoi tessellation and watershed algorithm, which has been shown to contain relevant cosmological information~\citep{kreisch_2022}.

For this reason, we explored a new methodology that takes advantage of the entire density field estimated through the Voronoi tessellation by \vide. 
For each Voronoi cell $i$ we assign a weight equal to their volume, $V_i$, multiplied by the mean tracer density at the corresponding redshift $z$, $n_{\rm g}(z)$: $w_i = V_i(z) \, n_{\rm g}(z)$. This quantity can be interpreted as the inverse of the normalized density of the Voronoi cell, $\rho_i/\langle\rho\rangle=1+\delta_i$. To avoid numerical noise in the tracer density, we fit the measured tracer density with a 4$^{\rm th}-$order polynomial in $z$, $n_{\rm g}(z)=\sum_{i=0}^4 a_i z^i$. 
The post-processing methodology consists of an iteration of the following steps. \\ 
i) We select the Voronoi cell of the watershed void characterized by the highest weight $w_{i_{\max}}$, i.e. the lowest density contrast, and measure the corresponding mean normalized density as $1+\Delta_{N_{\rm v}}= 1/w_{i_{\max}}$. 
We create a set of $N_{\rm v}=1$ elements, $I_{\rm v}=\{i_{\max}\}$, to which, step by step, we add the cells that build up the threshold void. \\
ii) We then identify the set of Voronoi cells adjacent to the selected one using the Delaunay scheme, i.e. the dual of the Voronoi tessellation~\citep{platen_2007,neyrinck_2008_zobov}, which we call $A_{\rm loop}$. We grow the threshold void by adding all the Voronoi cells, moving them from $A_{\rm loop}$ to $I_{\rm v}$, from the highest to the lowest weight. For each added cell, we update the iteration counter by one, $N_{\rm v} \rightarrow N_{\rm v} + 1$, and the corresponding mean normalized density as: 
\begin{equation}\label{eq:delta_v_alg}
1+\Delta_{N_{\rm v}} = \frac{\sum_{i \in I_{\rm v}}n_{\rm g}^{-1}(z_i)}{\sum_{i \in I_{\rm v}}V_i} \simeq \frac{N_{\rm v}n_{\rm g}^{-1}(z_{i_{\rm max}})}{\sum_{i \in I_{\rm v}}V_i} .
\end{equation}
The last approximate equality follows from the fact that the redshift extension of voids is negligible with respect to the variation scale of $n_{\rm g}(z)$, $\Delta z \ll 1$. \\
iii) If $\Delta_{N_{\rm v}}$ exceeds the void formation threshold $\delta_{\rm v}$ we interrupt the iteration.  Otherwise, we select the Voronoi cell characterized by the highest weight $w_i$ from $A_{\rm loop}$, and we repeat step ii), avoiding to select cells already belonging to the growing threshold void. \\
Once all the cells belonging to the threshold void are considered, we compute the standard void quantities provided by \vide, using the Voronoi cells information~\citep{sutter_2015_vide}. In particular in this work we consider the volume weighted barycenter, ${\bf X}_{\rm v} = \sum_{i \in I_{\rm v}} w_i {\bf x}_i$, where $I_{\rm v}$ is the set of Voronoi cells of the void, and the effective radius, i.e. the radius of the sphere with a volume equal to the sum of the Voronoi cells\footnote{We note that we use $R_{\rm eff}$ to indicate the effective radius of {\tt VIDE} voids and $R_{\rm v}$ to indicate the effective radius of voids when post-processed with the algorithm described in this Section.},
\begin{equation}\label{eq:Reff}
R_{\rm v} = \left[\frac{3}{4 \pi} \sum_{i \in I_{\rm v}} V_i \right]^{1/3}.
\end{equation} 
These quantities are computed considering the linearly interpolated volume fraction of the last Voronoi cell in such a way that the void formation threshold is exactly matched, i.e. $f$ such that $(N_{\rm v}-1+f)/(\sum_{j=1}^{N_{\rm v}-1} w_{I_{\rm p}[j]} + f w_{I_{\rm p}[N_{\rm v}]}) = 1+\delta_{\rm v}$, where $j$ are the indexes of the $I_{\rm v}$ element set.
This algorithm is publicly available in the {\tt vorothreshold}\footnote{\url{https://github.com/Giovanni-Verza/Voronoi_postprocess}} {\tt Python} package~\citep{vorothreshold_code}.

This algorithm is very similar to the watershed one~\citep{platen_2007,neyrinck_2008_zobov}, but instead of stopping at the watershed, it stops when a precise value of the mean normalized density is reached, making it specifically designed to optimize the VSF analysis. Moreover, it does not assume any symmetry in measuring the mean density contrast, making this algorithm independent of the cosmology adopted to measure distances. More precisely, the identification of the Voronoi cells belonging to each void is cosmology independent (at least in a neighborhood of the true cosmology), from which it follows that the angular position and redshift of the barycenter of the void do not depend on the assumed cosmology, while the inferred volume of voids changes according to the \citet[AP,][]{AP_1979} correction (see Section~\ref{subsec:vsf_analysis} and Appendix~\ref{appendix} for details). 
The algorithm is therefore more informative about the evolution of voids along cosmic history, conserving the void shape information. Additionally, the void center definition keeps the large-scale information about the low density of the environment.

\subsection{Void size function model}\label{subsec:vsf_model}

To theoretically describe the VSF, we adopt the model presented in~\citet{verza_2024}. This relies on the merging of the excursion-set framework~\citep{peacock_1990,bond_1991} and the theory of Lagrangian density peaks~\citep{bardeen_1986}, through an effective scale-dependent void formation barrier. This framework, following both the excursion-set and peak theory, describes the distribution of dark matter halos and voids in Lagrangian space, and maps the corresponding statistic in Eulerian space. We recall that the Lagrangian space is the initial density field linearly evolved up to the epoch of interest. In this context, ``initial'' means at redshift hight enough to be fully described by linear theory, and ``linearly evolved'' means that the global amplitude of the density contrast field is rescaled with the linear growth factor. The Eulerian space is the fully non-linear evolution of the density field at the epoch of interest. The mapping from Lagrangian to Eulerian space is performed by considering how voids (or halos) evolve.

As in the standard excursion-set model, a void with a Lagrangian radius $R$ is considered formed at the Lagrangian position $q$ if $R$ is the maximum smoothing scale at which the filtered Lagrangian density contrast field crosses the Lagrangian void formation threshold $\delta_{\rm v}^{\rm L}$, without having crossed the threshold for collapse on any larger scale. The main improvements with respect to the~\citet{SVdW_2004} multiplicity function can be summarized as follows: i) in the standard excursion-set framework, the position $q$ is random, while in the model considered here, this is a minimum in the density contrast field filtered at the scale $R$; ii) together with the cloud-in-cloud (or void-in-void) exclusion typical of the excursion-set, this model accounts for the peak-in-peak exclusion; iii) this model properly accounts for the smoothing of the density contrast field, which is reflected both in accounting for correlations of the density contrast field smoothed at different lengths, and considering the exact relation between the smoothing length and the Lagrangian void size. It should be noted that this last point solves the normalization problems of the \citet{SVdW_2004} multiplicity function in Eulerian space~\citep{jennings_2013}. As a final remark, the Roman HLSS survey will provide a dense redshift galaxy catalog, which is reflected in the possibility of exploring voids down to small scales. Therefore, even if the \citet{SVdW_2004} model has already been used successfully to analyze BOSS data~\citep{contarini_2023,contarini_2024}, its accuracy is not sufficient for the Nancy Grace Roman Space Telescope and dense stage IV galaxy surveys: modeling the VSF over a wide range of scales requires a robust theoretical model at all scales. In fact, the \citet{SVdW_2004} and \citet{jennings_2013} models cannot reproduce the VSF measured in the 2000 square degrees galaxy lightcone~\citep{zhai_2021} adopted here, while the \citet{verza_2024} model does. 

The main quantity in modeling the size distribution of halos and voids is the formation threshold. Halos are collapsed objects, so the Lagrangian threshold for halo formation is the Lagrangian density contrast corresponding to a full collapse in Eulerian space~\citep{bond_1991,sheth_mo_tormen_2001,lee_2010,pace_2010}. However, voids do not undergo an analogous formation event and continue to expand forever (in the single-stream regime). 
Since voids do not undergo shell-crossing on the scales relevant to Roman \citep[see e.g.,][]{biswas_2010,pisani_2015}, this ensures that in principle evolved voids can always be mapped in Lagrangian space and {\it vice versa}, considering a simple analytical map for the threshold, radius, and position~\citep{verza_2024}.
It follows that the observed void formation threshold (in Eulerian space) can be any negative value, chosen according to the features of the given survey, to which corresponds a unique Lagrangian counterpart value.~\citep{pisani_2015,verza_2019,vsf_euclid,verza_2024}, as long as the theoretical model is coherently computed.

\begin{table}[t]
\centering
\begin{tabular}{cccc}
\toprule
z & $\alpha$ & $\beta$ & $\gamma$  \\ 
\bottomrule
\noalign{\vspace{0.4em}}
1.06--1.22 & $0.0967_{-0.0168}^{+0.0132}$ & $0.110_{-0.016}^{+0.021}$ & $1.81_{-0.59}^{+0.25}$ \\ 
\noalign{\vspace{0.4em}}
1.22--1.44 & $0.179_{-0.007}^{+0.008}$ & $0.0660_{-0.005}^{+0.006}$ & $2.62_{-0.78}^{+0.70}$ \\ 
\noalign{\vspace{0.4em}}
1.44--1.83 & $0.225_{-0.004}^{+0.017}$ & $0.0493_{-0.005}^{+0.013}$ & $1.98_{-0.65}^{+1.05}$ \\ 
\noalign{\vspace{0.4em}}
\hline
\bottomrule
\end{tabular}
\caption{Maximum posterior distribution values and 1D 68\% CL interval for the moving barrier parameters  in the true cosmology, see Eq.~\eqref{eq:moving_b}. The first column lists the redshift bins range, the other columns list the corresponding $\alpha$, $\beta$, and $\gamma$ parameters, respectively.}
\label{tab:params_vsf}
\end{table}

In particular, we adopt the multiplicity function of Eq.~(15) in~\cite{verza_2024}: 

\begin{align}\label{eq:multipl_f}
f(S) &= \frac{e^{-B^2_S / 2S}}{\sqrt{2 \pi S}}  \left\llbracket  \sqrt{\frac{\Gamma_{\delta \delta}}{2 \pi S}} \exp \left[ -\frac{S}{2\Gamma_{\delta \delta}} \left(\frac{B_S}{2S} - B'_S \right)^2\right]\right. +  \\
& \left. \frac{1}{2} \left(\frac{B_S}{2S} - B'_S \right) \left\{{\rm erf}\left[\sqrt{\frac{S}{2\Gamma_{\delta \delta}}} \left(\frac{B_S}{2S} - B'_S \right)\right]+1\right\} \right\rrbracket , \nonumber
\end{align}
where 
\begin{equation}
S=\sigma^2(R)=\langle \delta_S^2 \rangle=\int \frac{{\rm d} k \, k^2}{2 \pi^2} P(k) |W(kR)|^2\,.
\end{equation}
with $\delta_S$ the linear density contrast field filtered at the scale $S=S(R)$, $P(k)$ the linear power spectrum, $W(kR)$ the top-hat filter function in Fourier space, $\Gamma_{\delta \delta} = SD_S - 1/4$, and $D_S=\langle ({\rm d} \delta_S / {\rm d}S)^2 \rangle$. The moving barrier $B_S=B(S,\delta_{\rm v}^{\rm L})$, is a function of the physical void formation barrier $\delta_{\rm v}^{\rm L}$, while $B'_S = {\rm d} B(S) / {\rm d} S$.\footnote{In this work we use $\delta_{\rm v}^{\rm L}$ to refer to the void formation threshold in Lagrangian space; $\delta_{\rm v}$ without any superscript refers to the non-linear value.} For the VSF analysis, all the above quantities are computed using the {\tt excursion\_set\_functions}\footnote{\url{https://github.com/Giovanni-Verza/excursion_set_functions}}~\citep
{ex_set_code} publicly available {\tt C++}/{\tt Python} package.

\subsection{Methodology}\label{subsec:vsf_methodology}

\begin{figure}[t]
\centering
\includegraphics[width=\linewidth]{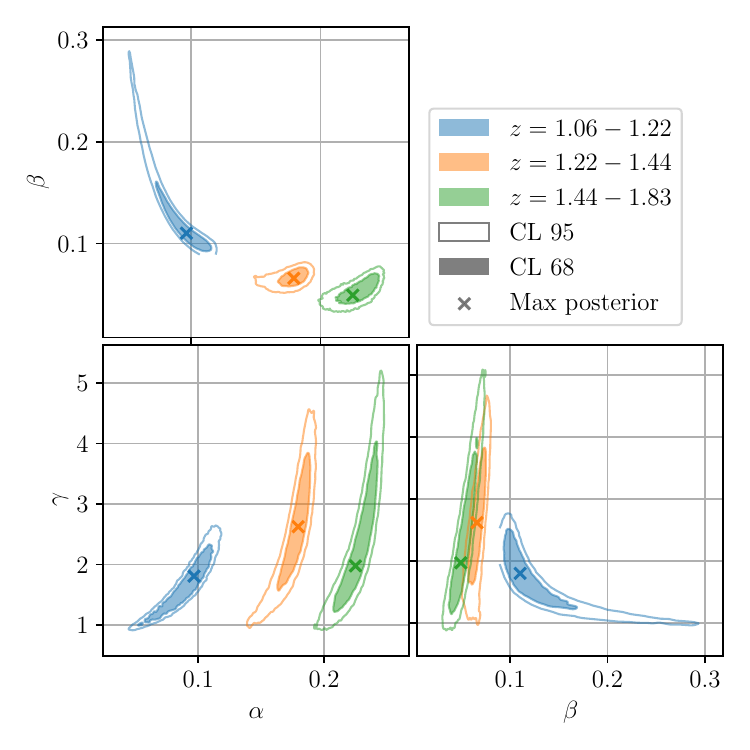}
\caption{2D marginalized posterior distributions for the moving barrier parameters in the true cosmology, Eq.~\eqref{eq:moving_b}, for each redshift bin, as labeled in the legend. The shaded area shows the 68\%CL, while the outer contours corresponds to 95\%CL. The crosses show the maximum of the posterior distributions.}
\label{fig:params_vsf}
\end{figure}

\begin{figure*}[t]
\centering
\includegraphics[width=0.9\linewidth]{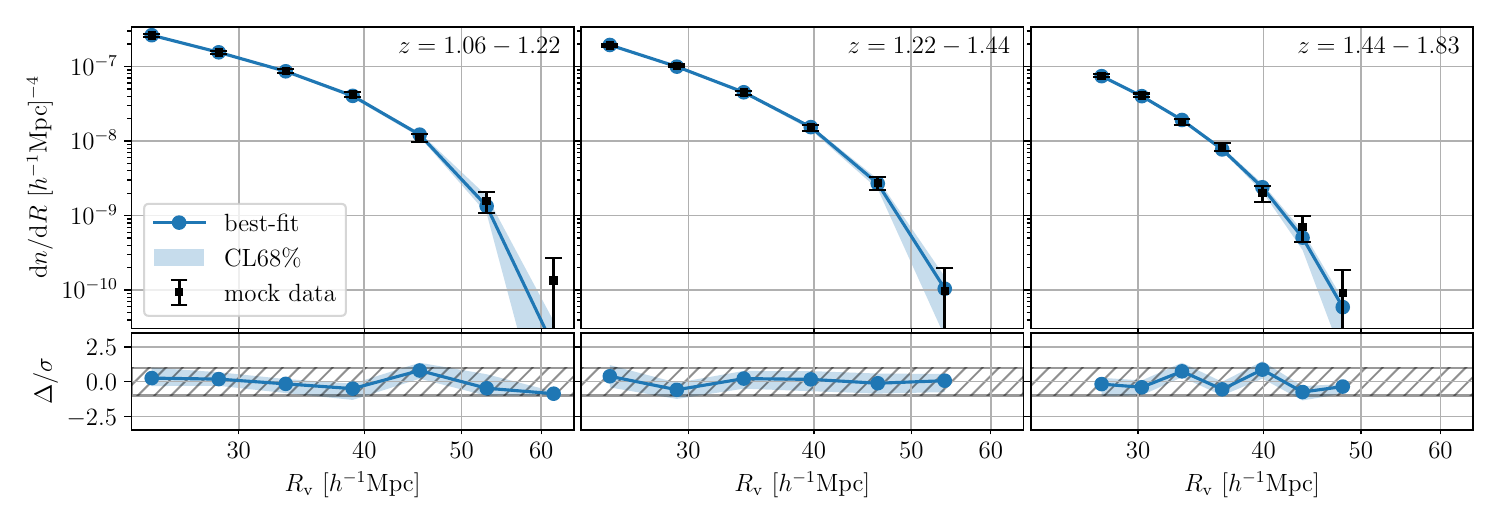}
\caption{From left to right: VSF from the Roman reference HLSS-like mock \citep{zhai_2021}, in each of the three redshift bins considered. Upper panels: black dots with error bars show the measured VSF from the post-processed void catalog as described in Section~\ref{subsec:vsf_cat} with $\delta_{\rm v,g}=-0.7$, the error bars are Poissonian; blue solid lines show the best-fit moving barrier calibrated in the true cosmology as described in Section~\ref{subsec:vsf_methodology}; blue shaded areas show 68\% CL. Lower panel: blue solid lines show the  relative values of the best VSF theoretical model with respect to measurements in $\sigma$ units; blue shaded areas show the 68\% CL; gray hatched areas show the $\pm 1\sigma$ interval.}
\label{fig:vsf_calibration}
\end{figure*}

The theoretical VSF model describes the number density of voids in the matter distribution, while in survey analyses we have access to the density contrast field of the galaxy distribution in redshift-space, that is with biased tracers. Voids are detected in this field, and therefore we have to take into account all the observational effects impacting the VSF. Some of these effects can be accounted for analytically. For example, galaxy bias in voids can be theoretically modeled, recovering the corresponding matter density contrast~\citep{verza_2022}. While the apparent enlargement of voids due to redshift-space distortions can be modeled, this neglects non-linear effects~\citep{correa_2021}. In fact it has been shown that even void detection can be affected~\citep{pisani_2015_peculiar_vel,correa_2022,massara_2022,radinovic_2024}, and the impact on constraints is a topic currently under investigation. Various other observational effects impact the mapping from the measured VSF in the observed galaxy distribution and the corresponding one in the matter distribution. For this reason, in this work we chose to model the effective model barrier for void formation to account for various important observational effects, such as: galaxy bias, linear and non-linear redshift-space distortion effects, non-trivial evolution from Lagrangian to Eulerian space\footnote{We note that in this approach, the mapping from Lagrangian to Eulerian voids is implicitly accounted for in the moving barrier modeling.}, selection effects, etc. 
The moving barrier follows to functional shape of $B_S=B(S)$~\citep{verza_2024},
\begin{equation}\label{eq:moving_b}
B(S) = \alpha \left[ 1 + (\beta / S)^{\gamma} \right].
\end{equation}
The three parameters of the VSF model, namely $\alpha$, $\beta$, and $\gamma$ in the above equation, have a precise dependence on the voids depth and are functions of redshift, as they depend on the linear growth factor and on the observed tracer population. To assess this dependence in the context of this analysis, we performed a Markov chain Monte Carlo (MCMC) against the measured VSF in the true cosmology. 
Figure~\ref{fig:params_vsf} shows the Markov chain Monte Carlo (MCMC) best-fit of the moving barrier parameters (crosses), with the 68\% and 95\% confidence level (CL) represented by the shaded areas and contours, respectively. Each color corresponds to a different redshift bin, as described in the legend. Table~\ref{tab:params_vsf} lists the corresponding best-fit values and 1D 68\% CL intervals. A clear redshift dependence of $\alpha$ and $\beta$ is visible, as expected, while $\gamma$ can be considered constant at the precision level allowed by simulated data. We note that, according to Eq.~\eqref{eq:moving_b}, $\alpha$ drives the overall height of the barrier, while $\beta$ determines the scale at which the exponential cut of the VSF occurs. 

It follows that, given the tracer evolution and the various observational effects we considered, the redshift dependence of these parameters is expected. 
Such expected redshift evolution is confirmed by the posterior distributions of the parameters used in this analysis in each redshift bin, shown in Figure~\ref{fig:params_vsf}. While in this work we characterize these parameters independently in each redshift bin considered (which in some way accounts for a redshift dependence), this does not correspond to having a model of the redshift evolution for these parameters. Developing such a model would require for example a unique relationship  depending on redshift for each parameter, and would effectively reduce the number of nuisance parameters for VSF analyses. This is, however, beyond the scope of this paper, and will be explored in a future work.
Alongside the redshift evolution, the effective barrier parameterization may be affected by dependencies on cosmology and tracer population. The strongest impact is expected to come from the tracer population via their bias. This would impact the global amplitude of the barrier but not the shape. A similar cosmological dependence is expected from those parameters that modify the amplitude of the linear matter power spectrum, without impacting its shape, as, e.g., $\sigma_8$ or a dynamical dark energy component. Finally, a minor dependence on the linear matter power spectrum shape is expected, which can introduce a cosmology dependence for those parameters that impact the power spectrum shape, such as $\Omega_{\rm m}$. This is due to minor changes in the shape of the correlation between the random walk steps at different scales, from which the multiplicity function is generated. This effect is, however, expected to be sub-dominant~\citep{verza_2024}. As mentioned above, exploring the redshift dependence for tracers and cosmology is beyond the scope of this work: we plan to fully characterize these dependencies in future works, by using lightcone mock realizations.

Figure~\ref{fig:vsf_calibration} shows the VSF in each of the three redshift bins considered. In the upper panels, black dots with error bars show the measured VSF from the post-processed void catalog as described in Section~\ref{subsec:vsf_cat} with $\delta_{\rm v,g}=-0.7$ the density contrast threshold used to build the void catalog for the VSF analysis in the galaxy distribution, the error bars are Poissonian. Blue solid lines show the best-fit moving barrier calibrated as described in Section~\ref{subsec:vsf_methodology}, while blue shaded areas show 68\% confidence level (CL). In the lower panel, blue solid lines show the relative values of the best VSF theoretical model with respect to measurements in $\sigma$ units, blue shaded areas show the 68\% CL, and gray hatched areas show the $\pm 1\sigma$ interval.

\subsection{Analysis and results}\label{subsec:vsf_analysis}

To study the constraining power of the VSF in the Roman reference HLSS, we perform a Bayesian analysis using a MCMC to sample the posterior distribution
\begin{equation}
{\cal P}(\pmb{\Theta} | {\cal D})  \propto {\cal L} ({\cal D} | \pmb{\Theta}) \, p(\pmb{\Theta})\,,
\end{equation}
where ${\cal L}({\cal D} | \pmb{\Theta})$ is the likelihood, ${\cal D}$ is the data vector, i.e. the measured VSF in the three redshift bins, $\pmb{\Theta}$ is the array of the cosmological parameters explored plus any extra model parameters, and $p(\pmb{\Theta})$ is the prior distribution of the parameters. We consider a Gaussian likelihood using the theoretical VSF with the calibration of the effective barrier parameters, presented in Section~\ref{subsec:vsf_model} and~\ref{subsec:vsf_methodology},
\begin{align}
\log \left[ {\cal L}({\cal D} | \pmb{\Theta}) \right] =& -\frac{1}{2} \sum_{ij} (n^{\cal D}_i - n^T_i) \Sigma^{-1}_{ij} (n^{\cal D}_j - n^T_j) \nonumber \\
& -\frac{N_{\cal D}}{2} \log (2 \pi )
-\frac{1}{2} \log (\det \Sigma ).
\end{align}
The indexes $i=i(i_r,i_z)$ and $j=j(j_r,j_z)$ run over all the radius bins, $i_r$, of the VSF and all the redshift bins considered, $i_z$, i.e. $i$ and $j$ run from 1 to $N_{\cal D}$, which is the length of the data vector considered. The $n^{\cal D}_i$ quantity refers to the number density of voids with radius in the radius bin $i_r$ and redshift bin $i_z$, $n^T_i$ refers to the corresponding theoretical  prediction, i.e. the integration of the theoretical VSF over the radius bin. Previous studies on mock catalogs showed that off-diagonal terms of the covariance matrix $\Sigma_{ij}$ of the measured VSF are subdominant or negligible~\citep{bayer_2021,kreisch_2022,pellicciari_2023,contarini_2023,contarini_2024,thiele_2024}. Therefore, aiming to produce cosmological forecasts, it is safe to consider a diagonal covariance, where the elements correspond to the Poissonian uncertainty provided by the number of voids in a given radius and redshift bin: $\Sigma_{ij} = \sigma_i^2 \delta^{\rm K}_{ij}$ where $\delta^{\rm K}_{ij}$ is the Kronecker delta, $\sigma_i = \sqrt{N_i}/V_{i_z}$, $N_i$ is the number of voids in the redshift bin $i_z$ with radius in the radius bin $i_r$, and $V_{i_z}$ is the volume of the redshift bin $i_z$. 
We note that we use here a Gaussian likelihood instead of a Poissonian one, since in previous works~\citep{vsf_euclid,beyond2pt} we did not find any relevant difference in the posterior distribution obtained with a Gaussian or a Poissonian prior, respectively.

It is important to note that to obtain the measured VSF we assumed a cosmological model, $\Lambda$CDM, with a fixed set of cosmological parameters, which we call ``fiducial cosmology''. This is because we converted the observed redshift of galaxies to comoving distances, and we ran the void finder and post-processed voids using these coordinates. At each step of the MCMC, we compute the posterior probability of an assumed ``true cosmology'', likely different from the ``fiducial'' one. It follows that not only the theoretical model, but also the inferred number density $n^{\cal D}_i$ and the corresponding covariance $\Sigma_{ij}$ change, due to the introduction of geometrical distortions: the \citet[AP,][]{AP_1979} and volume effects~\citep{vsf_euclid,verza_2023}.

The AP effect impacts the estimated void sizes and introduces an anisotropy between the orthogonal and the parallel directions with respect to the line-of-sight (LOS). Let us consider two points characterized by a mean redshift $z$, a difference in redshift $\Delta z \ll 1$ and an angular separation $\Delta \theta \ll 1 {\rm~rad}$. Their distance, decomposed in the radial and orthogonal components with respect to the LOS, is \citep{eisenstein_2005,xu_2013}:
\begin{align}\label{eq:AP_par_perp}
r^{\rm true}_\parallel &= \frac{H^{\rm fid}(z)}{H^{\rm true}(z)} \, r^{\rm fid}_\parallel = q_\parallel \, r^{\rm fid}_\parallel , \\
r^{\rm true}_\perp &= \frac{D_{\rm A}^{\rm true}(z)}{D_{\rm A}^{\rm fid}(z)} \, r^{\rm fid}_\perp = q_\perp \, r^{\rm fid}_\perp , \nonumber
\end{align}
where $H(z)$ is the Hubble parameter and $D_{\rm A}(z)$ the comoving (angular-diameter) distance. The superscripts indicate whether the corresponding quantities are computed considering the fiducial or assumed true cosmology in each MCMC step, respectively. We note that the volume of a Voronoi cell is modified as $V^{\rm true}_{\rm cell} = q_\parallel q^2_\perp V^{\rm fid}_{\rm cell}$. Therefore, by construction, following Eq.~\eqref{eq:Reff}, the void radius transforms as $R^{\rm true}_{\rm v}=q_\parallel^{1/3}q_\perp^{2/3} R^{\rm fid}_{\rm v}$. This result is already known for voids~\citep{hamaus_2020,correa_2021, vsf_euclid}, for which, however, this relation is an approximation and subject to the cosmology dependence on the void finder or post-processing procedure~\citep{correa_2021, radinovic_2024}. In our case, relying on the Voronoi tessellation, the relation is exact in the limit in which Eqs.~\eqref{eq:AP_par_perp} are satisfied~\citep{verza_2023}. See Appendix~\ref{appendix} for a detailed discussion and analysis on the impact of fiducial cosmology on the Voronoi tessellation and void finding procedure.

The volume effect changes the number density $n^{\cal D}_i$ and the associated uncertainty $\sigma_i$ due to the change in the inferred redshift bin volume, 
\begin{equation}
V_{i_z} = \frac{\Omega_{[{\rm rad}]}}{3} \left[D_{\rm A}^3(z_{\rm out}) - D_{\rm A}^3(z_{\rm in})\right]
\end{equation}
where $\Omega_{[{\rm rad}]}$ is the sky area in steradian, $D_{\rm A}(z)$ is the comoving distance corresponding to an object at redshift $z$, while $z_{\rm in}$ and $z_{\rm out}$ are the inner and outer redshift limits of the redshift bin $i_z$. It follows that
\begin{align}
\sigma^{\rm true}_i &= \sigma^{\rm fid}_i \frac{D_{\rm A, fid}^3(z_{\rm out}) - D_{\rm A, fid}^3(z_{\rm in})}{D_{\rm A, true}^3(z_{\rm out}) - D_{\rm A, true}^3(z_{\rm in})}, \\ 
n_i^{\rm true} &= n^{\rm fid}_i \frac{D_{\rm A, fid}^3(z_{\rm out}) - D_{\rm A, fid}^3(z_{\rm in})}{D_{\rm A, true}^3(z_{\rm out}) - D_{\rm A, true}^3(z_{\rm in})}.
\end{align}
The above relations assume that the number of voids is conserved, i.e. their detection and identification do not depend on the cosmology assumed in running the void finder and in post-processing the void catalog. This condition is only approximate for spherical threshold voids~\citep{correa_2021,radinovic_2024}, while it is exactly satisfied by both the void finder and the post-process procedure presented in this work (see Section~\ref{subsec:vsf_cat}). This is because, under a smooth change of the tracer distance from the observer, the Voronoi tessellation would provide a different cell volume, but the topological structure does not change. In particular, the identification of the Voronoi cells corresponding to the minima in the density field does not change, as well as the identification of the Voronoi cells building up the void (see Appendix~\ref{appendix}). In other words, the redshift and angular coordinates of the minima and void barycenter are not impacted by the assumed cosmology~\citep{verza_2023}. 
Moreover, the threshold value is not affected by geometrical distortions, due to cancellations of the AP terms. 
We note that $n^{\rm true}_{\rm g}(z) = q^{-1}_\parallel q^{-2}_\perp n^{\rm fid}_{\rm g}(z)$, while $V^{\rm true}_i = q_\parallel q^2_\perp V^{\rm fid}_i$. In the limit in which the AP correction is valid, $\sum_{i \in I_{\rm v}}V^{\rm true}_i = q_\parallel q^2_\perp \sum_{i \in I_{\rm v}} V^{\rm fid}_i$, where $q_\parallel$ and $q^2_\perp$ are evaluated at the redshift of the void center. From Eq.~\eqref{eq:delta_v_alg} it follows $\delta^{\rm fid}_{\rm v} = \delta^{\rm true}_{\rm v}$. Therefore, both the relative minima and the cells belonging to threshold voids are conserved, i.e. the void catalog is cosmology independent (see Appendix~\ref{appendix}). We note that even this last relationship is exact for our void catalog, while it is approximate for spherical voids.

\begin{figure*}[t!]
\centering
\includegraphics[width=0.85\linewidth]{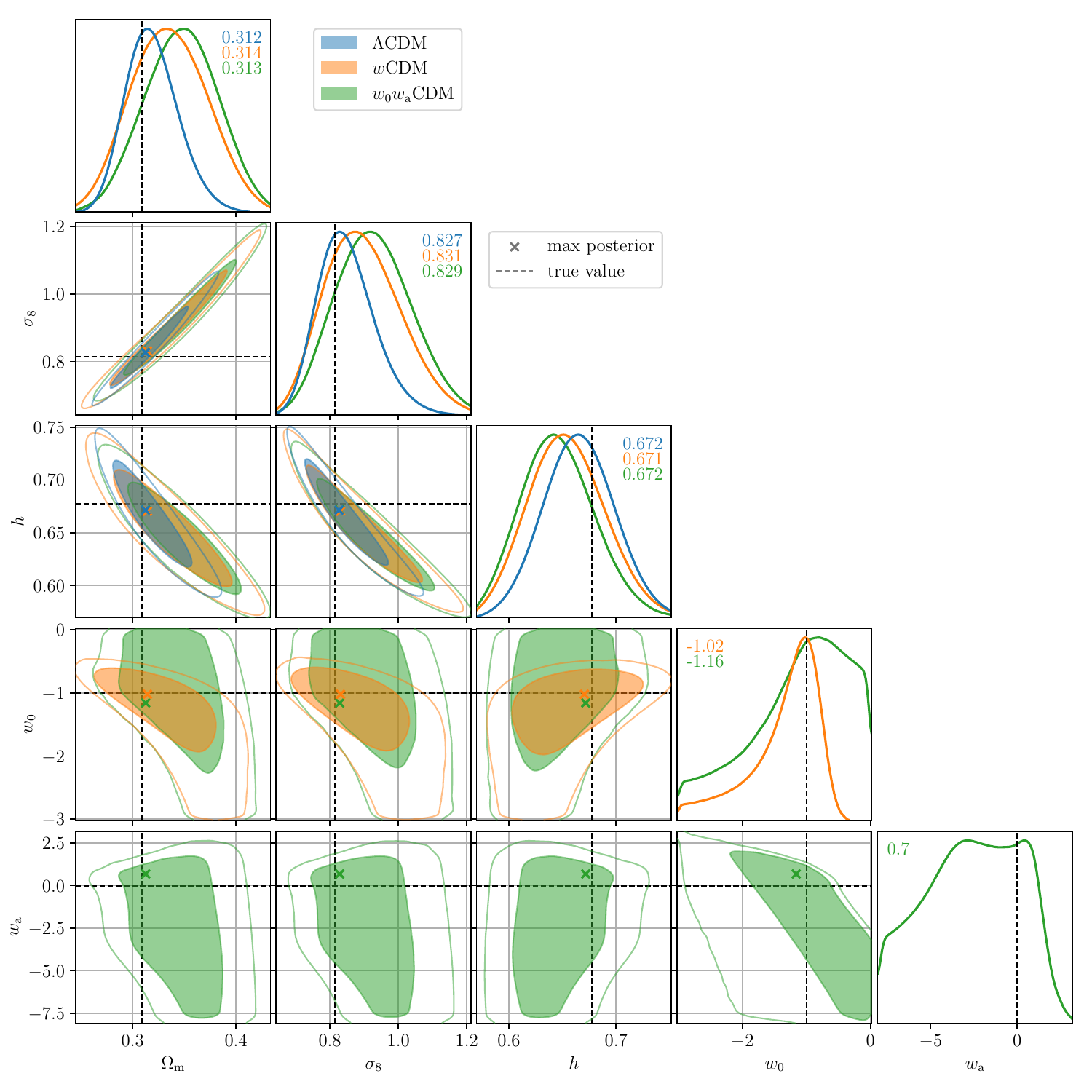}
\caption{VSF: posterior distributions of the cosmological parameters explored in the optimistic scenario, for the three cosmological models explored: $\Lambda$CDM (blue), $w$CDM (orange), $w_0 w_{\rm a}$CDM (green). The filled internal region shows the 68\% CL, the outer line shows the 95\% CL. Black dashed lines show the true values, crosses correspond to the maximum of the posterior distribution, the numbers in the panels along the diagonal list the parameter values corresponding to the maximum of the likelihood.}
\label{fig:results_vsf_optimistic}
\end{figure*}

\begin{figure*}[t!]
\centering
\includegraphics[width=0.85\linewidth]{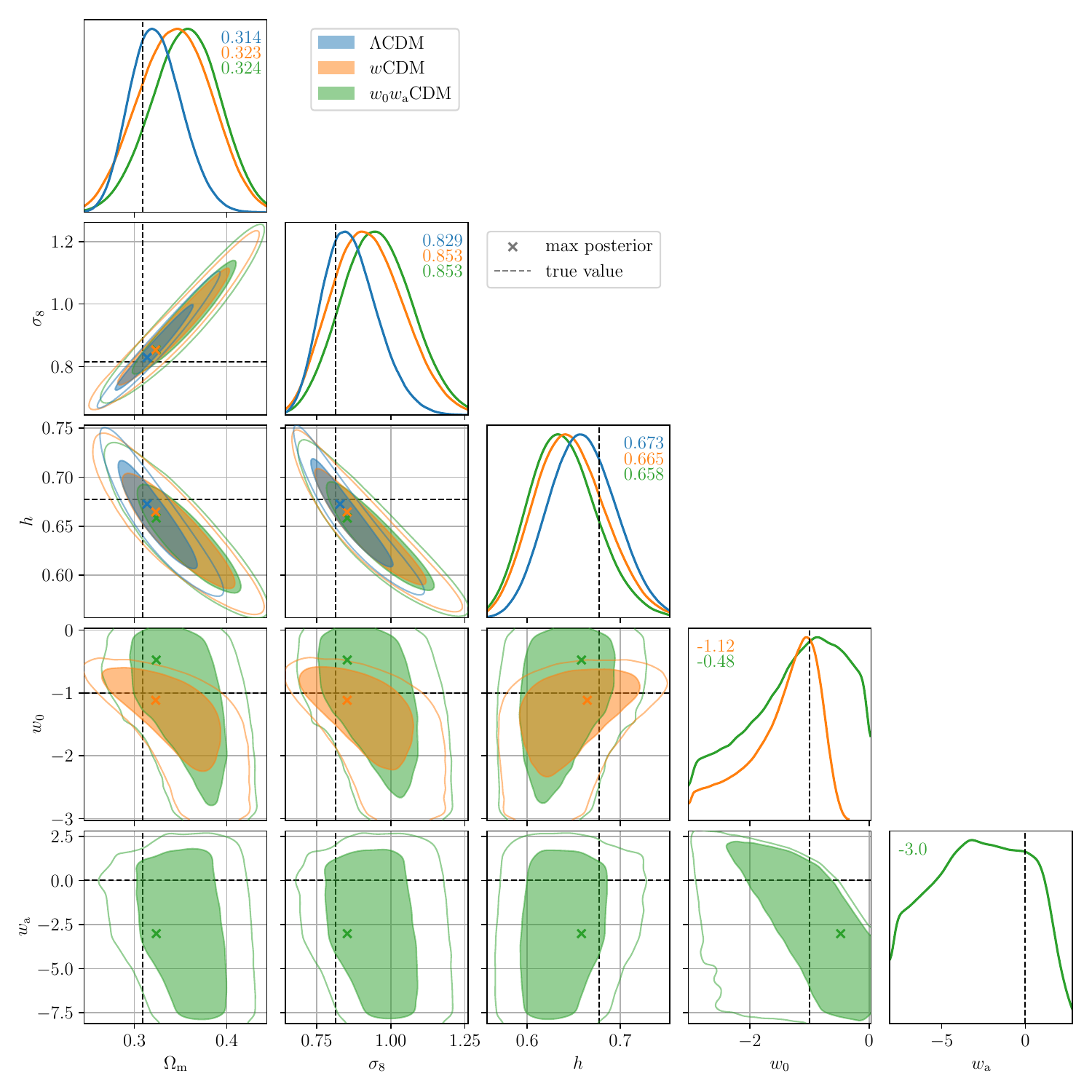}
\caption{VSF: posterior distributions of the cosmological parameters explored in the pessimistic scenario, for the three explored cosmological models. The plot is organized as Figure~\ref{fig:results_vsf_optimistic}.}
\label{fig:results_vsf_pessimistic}
\end{figure*}

\begin{table*}
\centering
\begin{tabular}{ccccccc}
\toprule
\noalign{\vspace{0.075cm}}
Model & & $\Omega_{\rm m}$ & $\sigma_8$ & $h$ & $w_0$ & $w_{\rm a}$  \\ 
\noalign{\vspace{0.1cm}}
\toprule
\noalign{\vspace{0.15cm}}
True & & $0.3089$ & $0.8147$ & $0.6774$ & $-1$ & $0$ \\ 
\noalign{\vspace{0.4em}}
\cline{1-7}
\noalign{\vspace{0.4em}}
\multirow{3}{*}{$\Lambda$CDM} & opt. & $0.3125^{+0.0283}_{-0.0216}$ & $0.8268^{+0.0880}_{-0.0714}$ & $0.6715^{+0.0263}_{-0.0395}$ & $-1$ & $0$  \\ 
\noalign{\vspace{0.4em}}
\cline{3-7}
\noalign{\vspace{0.4em}}
& pess. & $0.3138^{+0.0352}_{-0.0203}$ & $0.8288^{+0.1126}_{-0.0675}$ & $0.6726^{+0.0214}_{-0.0505}$ & $-1$ & $0$ \\ 
\noalign{\vspace{0.4em}}
\cline{1-7}
\noalign{\vspace{0.4em}}
\multirow{3}{*}{$w$CDM} & opt. & $0.3142^{+0.0571}_{-0.0175}$ & $0.8311^{+0.1610}_{-0.0589}$ & $0.6705^{+0.0169}_{-0.0544}$ & $-1.022^{+0.318}_{-0.542}$ & $0$ \\ 
\noalign{\vspace{0.4em}}
\cline{3-7}
\noalign{\vspace{0.4em}}
& pess. & $0.3230^{+0.0614}_{-0.0174}$ & $0.8527^{+0.1820}_{-0.0587}$ & $0.6645^{+0.0164}_{-0.0594}$ & $-1.117^{+0.425}_{-0.613}$ & $0$ \\ 
\noalign{\vspace{0.4em}} 
\cline{1-7}
\noalign{\vspace{0.4em}}
\multirow{3}{*}{$w_0 w_{\rm a}$CDM} & opt. & $0.3126^{+0.0703}_{-0.0008}$ & $0.8288^{+0.1998}_{-0.0218}$ & $0.6717^{+0.0052}_{-0.0627}$ & $-1.160^{+1.160}_{-0.225}$ & $0.69^{+0.47}_{-5.28}$ \\ 
\noalign{\vspace{0.4em}}
\cline{3-7}
\noalign{\vspace{0.4em}}
& pess. & $0.3237^{+0.0696}_{-0.0029}$ & $0.8533^{+0.2134}_{-0.0210}$ & $0.6583^{+0.0121}_{-0.0593}$ & $-0.476^{+0.476}_{-1.091}$ & $-3.02^{+4.20}_{-1.92}$ \\  
\noalign{\vspace{0.15cm}}
\hline
\bottomrule
\end{tabular}
\caption{VSF: maximum posterior distribution values and 1D 68\% CL interval for the cosmological parameters explored, for each considered cosmological model (we show both the optimistic and pessimistic scenarios).}
\label{tab:vsf_constraints}
\end{table*}

In this work, we explore three different cosmological models. The first is the flat $\Lambda$CDM model, where the cosmological parameters explored are $\pmb{\Theta}_{\Lambda {\rm CDM}} = \{\Omega_{\rm m}, \sigma_8, h \}$, while $n_{\rm s}$ is kept fixed to the value used in the simulation, see Section~\ref{sec:simu_vide}. We note that $\sigma_8$, the normalization of the linear power spectrum at $z=0$, has a one to one mapping with $A_{\rm s}$, the amplitude of scalar fluctuations after inflation. In the MCMC, we explore $A_{\rm s}$ and obtain $\sigma_8$ as a derived parameter. We use wide flat priors for all parameters, $\Omega_{\rm m} \in (0.15,0.65)$, $A_{\rm s} \in (10^{-10},5\times 10^{-9})$, $h \in (0.3,2)$. The second model is the $w$CDM model, an extension of the $\Lambda$CDM model with a dark energy component characterized by a constant equation of state, $w$. The parameters explored are $\pmb{\Theta}_{w{\rm CDM}} = \pmb{\Theta}_{\Lambda {\rm CDM}} \cup \{ w \}$. The prior distributions for the parameters in common with $\Lambda$CDM are the same as the ones adopted for $\Lambda$CDM. For $w$ we consider a flat prior with the range $w \in (-3,0)$. The third and last model we consider is a $w_0 w_{\rm a}$CDM model, i.e. we consider a dynamical dark energy component, characterized by the Chevallier--Polarski--Linder (CPL) equation of state \citep{chevallier_polarski_2001, linder_2003}, which is the Taylor expansion of any possible equation of state truncated at the linear order,
\begin{equation}
\label{eq: w0wa}
w(a) = w_0 + (1-a) \, w_{\rm a} \ \Rightarrow \ w(z) = w_0 + \frac{z}{1+z} w_{\rm a} .
\end{equation}
The parameters explored are $\pmb{\Theta}_{w_0 w_{\rm a}{\rm CDM}} = \pmb{\Theta}_{\Lambda {\rm CDM}} \cup \{w_0, w_{\rm a} \}$. The priors for $\pmb{\Theta}_{\Lambda {\rm CDM}}$ parameters are as in the $\Lambda$CDM case. We use flat priors for both $w_0$ and $w_{\rm a}$, with $w_0 \in (-3,0)$, $w_{\rm a} \in (-8,5)$.

For each cosmological model, we perform two analyses, an optimistic one, in which the moving barrier parameters of Section~\ref{subsec:vsf_methodology} are considered fixed to the best fit value, and a pessimistic one, in which the effective barrier parameters are considered as nuisance parameters, using as prior for $\alpha$, $\beta$, and $\gamma$ of Eq.~\eqref{eq:moving_b} of each redshift bin the posterior of the calibration described in Section~\ref{subsec:vsf_methodology}, resulting in a total of 9 nuisance parameters. This choice is due to the fact that we want to explore the constraining power coming from a VSF analysis performed on Roman-like data. The optimistic scenario represents the case in which the theoretical model is fully understood and is able to provide a robust prediction; while the pessimistic case corresponds to the case in which the dependence of the moving barrier on the tracer distribution is unknown.
An intermediate step among these two cases would consist of modeling the redshift dependence of the effective barrier parameters. This would considerably decrease the number of nuisance parameters; its implementation is however beyond the scope of this work. The MCMC analyses are performed using the {\tt emcee}\footnote{\url{https://github.com/dfm/emcee}}~\citep{emcee} Python package, considering 48 walkers. Each chain is post-processed by removing the first 2000 steps. Moreover, to ensure that the different steps are not correlated, we consider 1 chain step every 15 computed, resulting in a final chain length of around $2 \times 10^7$, for each of the models considered.

Figures~\ref{fig:results_vsf_optimistic} and~\ref{fig:results_vsf_pessimistic} show the 2D  marginalized posterior distributions of the explored cosmological models, $\Lambda$CDM (blue), $w$CDM (orange), and $w_0 w_{\rm a}$CDM (green). Shaded areas show the 68\% CL, the outer lines show the 95\% CL, crosses correspond to the maximum of the posterior distribution, and dashed gray lines to the true values of the parameters. 
Table~\ref{tab:vsf_constraints} lists the best-fit values and 1D 68\% CL of each cosmological parameter in both the optimistic and pessimistic cases for each cosmological model considered. 
It can be noted that in the pessimistic case the 68\% CL of each cosmological parameter is slightly enlarged. This is due to the nuisance parameters that increase the dimensionality of the posterior. It is important to recall that in a more realistic case, when the effective barrier parameters will be modeled as a function of redshift, the number of nuisance parameters will decrease with respect to the one explored here.

To conclude this Section, we want to stress that this analysis considers the void size function alone. A great advantage of the VSF and other cosmic void statistics comes from their power when combined with other probes~\citep{pisani_2015,bayer_2021,kreisch_2022,pellicciari_2023,kreisch_2022, vsf_euclid}. Indeed it is possible to greatly tighten the constraints by combining void statistics among themselves~\citep[see e.g.,][and discussion in Section~\ref{sec:conclu}]{kreisch_2022, vsf_euclid}, and with galaxy statistics such as cluster counts and galaxy two-point statistics~\citep{bayer_2021,pellicciari_2023,vsf_euclid}. The additional power is due to various reasons. Firstly, cosmic voids probe different scales with respect to clusters and galaxies~\citep{pellicciari_2023,contarini_2023}. Secondly, since voids have an extended size, the AP effect on voids acts in a different way compared to how it acts on galaxy statistics. 
Finally, galaxies and voids probe different environments. Interestingly, the underdense environment of voids is representative of less evolved and pristine regions in the Universe~\citep{lavaux_2012,bos_2012,cautun_2014,pisani_2015,verza_2022}. Also, due to their underdense nature, voids are an environment where the dark energy-dark matter density ratio is higher than in the mean universe~\citep{cautun_2014,pisani_2015,verza_2019}.

Along with the VSF, the other void statistic we consider is the VGCF, discussed in the following Section.

\section{The void-galaxy cross-correlation function}\label{sec:vgcf}

\subsection{Theory}
\label{subsec: theory ccf}
The void-galaxy cross-correlation function (VGCF), $\xi(r)$, quantifies the excess probability of finding a galaxy at a comoving distance $r$ from the center of a void, compared to what would be expected in a random, uncorrelated galaxy distribution. In essence, $\xi(r)$ measures how much more (or less) likely it is to find galaxies near void centers relative to a uniform galaxy distribution across the Universe, making it a valuable statistical tool for studying average void shapes. While individual voids exhibit a variety of shapes and sizes, by stacking voids and analyzing the resulting average void shape, we expect—based on the cosmological principle of homogeneity and isotropy—that the average void is statistically spherical in real space.
Observed voids show distortions, arising when converting the measured quantities, i.e., redshifts and angular coordinates, into distances. Hence the observed stacked void profile deviates from spherical symmetry. This symmetry is disrupted by geometrical distortions, such as AP distortions, which arise from the use of an incorrect cosmological model when converting from redshifts and sky coordinates to distances, and redshift space distortions (RSD), which are caused by Doppler shifts due to the peculiar velocities of galaxies along the line of sight. 
Through the modelling of such distortions, the VGCF is widely used in void cosmology as a tool for constraining cosmological parameters~\citep[e.g.,][]{hamaus_2017, hamaus_2020, nadathur_2020, radinovic_2023, correa_2022, fraser_2024}.  
Specifically, geometric distortions enable the so-called AP test. This test relies on the principle that stacked voids will only appear spherically symmetric if the fiducial cosmological model used to convert redshifts into distances is correct~\citep{ryden_1995, lavaux_2012}.  To perform a successful AP test, it is essential to account for the presence of RSD, which requires careful modeling of peculiar velocities. These velocities can be modeled analytically by adopting a linear approximation~\citep{peebles_1980, kaiser_1987}, as described below, or can be modeled numerically by employing reconstruction algorithms \citep[see][]{degni_2025}.
For this scope, cosmic voids provide an optimal environment. In fact other cosmic web structures, such as galaxy clusters, filaments, and walls, underwent shell-crossing in early stages of their evolution.
This breaks both the validity of linear theory, and the Lagrangian fluid trajectories on which the above mentioned numerical methods rely. On the other hand cosmic voids are in the mildly non-linear regime, evolving in the single stream regime, reaching only mild or no shell-crossing in their inner regions~\citep{shandarin_2011, abel_2012, sutter_2014, hahn_2015,schuster_2023,schuster_2024}.

The relative peculiar velocity field $\textbf{u}$ with respect to the void center is the source of RSDs. Since the integrated density profile of a void $\Delta(r)$ obeys spherical symmetry,
\begin{equation}
    \label{eq: Delta(r)}
    \Delta(r)=\frac{3}{r^3}\int_0^r\delta(r')r'^2\mathrm{d}r',
\end{equation}
the relative velocity field $\textbf{u}$ is robustly described by linear theory \citep{peebles_1980}
\begin{equation}
    \label{eq: u Peebles}
    \mathbf{u}=-\frac{f(z)}{3}\frac{H(z)}{1+z}\Delta(r)\textbf{r},
\end{equation}
where $\mathbf{r}$ is the comoving distance from the void center (in real space) and $\delta$ the matter density contrast for the spherically symmetric stacked void.
Work with high-resolution N-body simulations shows that this spherical and linear approximation can accurately describe the velocity profile of stacked voids from large to very small scales, of the order of $\sim 1 h^{-1}$Mpc~\citep{hamaus_2014_profile,schuster_2023,schuster_2024}. This is likely due to the single-stream regime characterizing voids' evolution. In fact, according to the Zel'dovich approximation \citep{zeldovich_1970} and the Lagrangian perturbation theory, the linear Eulerian velocity field, Eq.~\eqref{eq: u Peebles}, well describes the fluid displacement field even beyond the breaking point of the Eulerian linear perturbation theory.

The observed galaxy redshift is a composition of the cosmological redshift $z$ and the Doppler shift sourced by galaxy peculiar velocities. It follows that converting the observed redshifts to comoving distances introduces a distortion along the line-of-sight (LOS). In the limit where linear theory is valid and the velocity component along the LOS satisfy ${\bf u}_\parallel/c \ll 1$, the comoving distance vector in redshift space, $\mathbf{s}$, reads
\begin{equation}
    \label{eq: mapping s(r)}
    \mathbf{s}=\mathbf{r}+\frac{1+z}{H(z)}\mathbf{u}_\parallel = \mathbf{r}-\frac{f}{3}\Delta(r)\mathbf{r}_\parallel.
\end{equation}
The last equality comes from the linear velocity field, Eq. \eqref{eq: u Peebles}. 
The relation between the real- and redshift-space VGCF can be derived considering the Jacobian of the real- to redshift-space mapping, Eq.~\eqref{eq: mapping s(r)},
\begin{equation}
    \label{eq: xis Cai}
    \xi^s(\mathbf{s})\simeq \xi(r)+\frac{f}{3}\Delta(r)+f\mu^2[\delta(r)-\Delta(r)],
\end{equation}
where redshift-space quantities are labeled with the superscript $s$, and  $\mu=r_\parallel/r$ is the cosine of the angle between the separation vector and the LOS \citep{cai_2016,hamaus_2017, hamaus_2020,hamaus_2022}.
Assuming a linear bias relation that connects the observed galaxy density field to the underlying matter density field, we obtain a relation of the form $\xi(r)=b\delta(r)$, and consequently also $\bar{\xi}(r)=b\Delta(r)$ with $\bar{\xi}(r)$ being the average VGCF in a sphere of radius $r$. It follows that Eq. \eqref{eq: xis Cai} becomes:
\begin{equation}
    \label{eq: xi model}
    \xi^s(\mathbf{s})\simeq \xi(r) + \frac{\beta}{3}\bar{\xi}(r)+\beta\mu^2[\xi(r)-\bar{\xi}(r)],
\end{equation}
where $\beta=f/b$. 
Indeed, while generically the void-galaxy bias term $b$ is not guaranteed to correspond to the standard galaxy large-scale effective bias, it is linearly related to it and converges to this value for sufficiently large voids~\citep{sutter_2014,pollina_2017,verza_2022}. This equation, together with Eq.~\eqref{eq: mapping s(r)}, represents the theoretical model of the observed VGCF in redshift-space, according to linear theory. We note that the real-space quantities appearing in Eq.~\eqref{eq: xi model}, i.e. $\xi(r)$ and $\bar{\xi}(r)$, are not directly observable. 
Since the real-space quantities are not known, we rely on the deprojection technique on the VGCF integrated along the LOS in redshift space, $\xi^s_p(s_\perp)$ \citep{pisani_2014, hawken_2017}. 
The $\xi^s_p(s_\perp)$ is identical to its real-space counterpart $\xi_p(r_\perp)$, as $s_\perp = r_\perp$ and both are obtained by  marginalizing over the LOS component, the one affected by RSD. 
So, with the forward Abel transform \citep{abel_1842,bracewell_1999} we obtain $\xi^s_p(s_\perp)$ from $\xi^s(\mathbf{s})$: 

\begin{equation}
    \label{eq: deprojection 2}
    \xi^s_p(s_\perp)=\int \xi^s(\mathbf{s}) \mathrm{d}s_\parallel=2\int_{s_\perp}^\infty r \xi(r) \left(r^2-s_\perp^2\right)^{-1/2}\mathrm{d}s_\perp.
\end{equation}
Then the inverse Abel transform \citep{abel_1842, bracewell_1999} provides the inverse mapping from $\xi^s_p(s_\perp)$ to get $\xi(r)$. This technique effectively allows to map $\xi^s_p(s_\perp)$ into the 3D spherically symmetric VGCF in real-space: 
\begin{equation}
    \label{eq: xi deprojection}
    \xi(r)=-\frac{1}{\pi}\int_r^\infty\frac{\mathrm{d}\xi^s_p(s_\perp)}{\mathrm{d}s_\perp}\left(s_\perp^2-r^2\right)^{-1/2}\mathrm{d}s_\perp.
\end{equation}

Following \cite{hamaus_2020, hamaus_2022}, we adopted an empirically motivated modification of the model presented in Eq.~\eqref{eq: xi model}, considering two nuisance parameters, ${\cal M}$ and ${\cal Q}$. The modified version of the model is as follows:
\begin{equation}
    \label{eq: xi model M Q}
    \xi^s(\mathbf{s})= \mathcal{M}\biggl\{ \xi(r) + \beta\bar{\xi}(r)+2\mathcal{Q}\beta\mu^2[\xi(r)-\bar{\xi}(r)]\biggr\},
\end{equation}
and the mapping between coordinates in real and redshift space becomes:
\begin{equation}
    \label{eq: mapping r s M}
    r_\perp=s_\perp \quad r_\parallel=s_\parallel\left[ 1-\mathcal{M}\frac{\beta}{3}\bar{\xi}(r)\right]^{-1}.
\end{equation}
The parameter $\mathcal{M}$, a monopole-like term, acts as a free amplitude of the deprojected correlation function $\xi(r)$ in real space, compensating for potential biases introduced by the deprojection method and the possible contamination of the void sample by random Poisson fluctuations, which can suppress the monopole and quadrupole amplitudes \citep{cousinou_2019}. On the other hand, the parameter $\mathcal{Q}$, the quadrupole-like term, corrects for selection effects that can arise when voids are identified in the anisotropic redshift space \citep{pisani_2015_peculiar_vel, correa_2021, correa_2022}. 
These effects, including shell-crossing and the virialization process, impact the structure of void boundaries in redshift space \citep{hahn_2015}, leading to the familiar Finger-of-God (FoG) effect \citep{jackson_1972}. These distortions may influence the Jacobian terms in Eq. \eqref{eq: xi model} and are part of the systematic effects that should be explored with further detail in future works.

So far, we have addressed the modeling of the VGCF in the presence of RSD alone. The AP effect, a geometric distortion, is modeled by expressing the fiducial coordinates, $r^\mathrm{fid}$, in terms of the true coordinates, $r^\mathrm{true}$. This coordinate transformation is applied using Eq. \eqref{eq:AP_par_perp}, which introduces the quantities $q_\parallel$ and $q_\perp$, 
\begin{equation}
    \label{eq: mapping r s M AP}
    r_\perp=q_\perp s_\perp \quad r_\parallel=q_\parallel s_\parallel\left[ 1-\mathcal{M}\frac{\beta}{3}\bar{\xi}(r)\right]^{-1}.
\end{equation}
From this equation, we can obtain $r=\sqrt{r^2_\parallel+r^2_\perp}$ and $\mu=r_\parallel/r$, entering in Eq.~\eqref{eq: xi model M Q}. As $r_\parallel$ itself explicitly depends on the quantities we want to derive, it's possible to evaluate them by iteration of Eq.~\eqref{eq: mapping r s M AP} \citep{hamaus_2020,hamaus_2022}. The combination of the $q_\parallel$ and $q_\perp$ parameters can be inferred from the eccentricity of the objects.
To quantify this, it is useful to define the parameter linked to AP distortions, $\epsilon$, as:
\begin{equation}
    \label{eq: epsilon}
    \epsilon\equiv\frac{q_\perp}{q_\parallel}=\frac{D_\mathrm{A}^\mathrm{true}(z)H^\mathrm{true}(z)}{D_\mathrm{A}^\mathrm{fid}(z)H^\mathrm{fid}(z)}.
\end{equation}
If the fiducial cosmology matches the true one, then $\epsilon = 1$, and consequently, $r^{\rm fid}_\parallel = r^{\rm true}_\parallel$ and $r^{\rm fid}_\perp = r^{\rm true}_\perp$. Conversely, if a measurement of $\epsilon$ differs from 1, it indicates that the chosen fiducial cosmology deviates from the true one. In this scenario, we can use the parameter $\epsilon$ to constrain $D_\mathrm{A}(z)$$H(z)$. 

Following the VSF case, for this analysis we adopt three different cosmological models: $\Lambda$CDM, $w$CDM, and $w_0w_\mathrm{a}$CDM. These models serve as our fiducial frameworks. The angular diameter distance, $D_\mathrm{A}(z)$, and the Hubble parameter, $H(z)$, can be generalized with the following expressions (in the case of a flat universe):
\begin{equation}
    \label{eq: DA(z)}
    D_\mathrm{A}(z) = \int_0^z \frac{c}{H(z')} \, \mathrm{d}z',
\end{equation}
\begin{equation}
    \label{eq: H(z)}
    H(z) = H_0 \sqrt{\Omega_\mathrm{m}(1+z)^3 + \Omega_\mathrm{de}(1+z)^{3(1+w_0+w_\mathrm{a})} F(z)},
\end{equation}
where 
\begin{equation}
F(z) = e^{-3w_\mathrm{a}z/(1+z)},
\end{equation}
$H_0$ is the present-day Hubble constant, $\Omega_\mathrm{m}$ is the matter density parameter, and $\Omega_\mathrm{de}$ is the dark energy density parameter (with $\Omega_\mathrm{de} \equiv \Omega_\Lambda$ in the $\Lambda$CDM model). The parameters $w_0$ and $w_\mathrm{a}$ describe the dark energy equation of state.

\subsection{Likelihood analysis}
\label{subsec: likelihood analysis}

In this sub-section we present measurements and the likelihood analysis of the VGCF. Those are performed using the {\tt Voiager}\footnote{\url{https://voiager.readthedocs.io}} publicly available package, which provides a pipeline to perform cosmological analyses using voids identified in large-scale structure survey data. This code measures dynamic and geometric shape distortions in void stacks and propagates the measurement down to constraints on cosmological parameters using Bayesian inference.

The data vector is represented by the VGCF measured in redshift space. As thoroughly discussed is Section~\ref{subsec: theory ccf}, this function is anisotropic along the LOS, and is therefore two-dimensional, $\xi^s(s_\perp,s_\parallel)$. By re-writing the $s_\perp$ and $s_\parallel$ quantities in term of $s$ and $\mu_s=s_\parallel/s$, it is possible to decompose the VGCF into multipoles of the Legendre polynomials $\mathcal{P}_\ell$ of order $\ell$:
\begin{equation}
\label{eq: multipoles ccf}
    \xi^s_\ell(s)=\frac{2\ell+1}{2}\int_{-1}^{1}\xi^s(s,\mu_s)\mathcal{P}_\ell(\mu_s)\mathrm{d}\mu_s .
\end{equation} 
The monopole, $\xi_0$, represents the average galaxy density profile within the void region.
The quadrupole, $\xi_2$, is identically null for statistically isotropic samples. Any deviation from zero indicates anisotropy: a positive quadrupole means the void density profile is compressed along the LOS, while a negative quadrupole suggests elongation along the LOS.
Similarly, a non-zero hexadecapole, $\xi_4$, indicates a breakdown of statistical isotropy. We note, however, that in linear theory this quantity is expected to be identically null, Eq.~\eqref{eq: xi model}, therefore its interpretation in terms of distortions of the void shape involves higher-order effects.

Even if the multipole decomposition provides a clear physical interpretation, actually for VGCF analyses it is possible to rely on either the 2D cross-correlation function $\xi^s(s_\perp,s_\parallel)$ for model fitting with coordinates along and perpendicular to the LOS (this will, of course, include information on both RSD and AP), or on the decomposition into multipoles of the Legendre polynomials. Here we use the 2D cross-correlation function, since this provides a better balance for the number of bins that sample the inner core of voids with respect to the bins sampling the slope of the void profile (inner bins are more strongly impacted by noise in e.g., the deprojection), but we also provide the multipoles corresponding to $\ell=0,2,4$, for completeness.

For our estimation of the VGCF  $\xi^\mathrm{s}(s_\perp,s_\parallel)$ we use the Landy-Szalay estimator \citep{landy_szalay}:
\begin{equation}
\label{eq: ccf LS}
    \xi^s(s_\perp,s_\parallel)=\dfrac{\langle \mathcal{D}_\mathrm{v}\mathcal{D}_\mathrm{g}\rangle -\langle \mathcal{D}_\mathrm{v}\mathcal{R}_\mathrm{g}\rangle - \langle \mathcal{R}_\mathrm{v}\mathcal{D}_\mathrm{g}\rangle +\langle \mathcal{R}_\mathrm{v}\mathcal{R}_\mathrm{g}\rangle}{\langle \mathcal{R}_\mathrm{v}\mathcal{R}_\mathrm{g}\rangle}.
\end{equation}
The angled brackets denote normalized pair counts of void-center v and galaxy, g, positions. The letter $\mathcal{D}$ refers to the data while $\mathcal{R}$ refers to the random positions. The pair counts are binned in $s_\perp$ and $s_\parallel$. We implemented a fixed binning scheme based on the effective void radius for each individual void, expressing all distances in units of the void radius $R_\mathrm{eff}$. 
Normalizing the void-galaxy distance with respect to the void radius is crucial to coherently overlap regions of similar density, and to enhance the topology of voids, such as the compensation wall, making our observable ideal to extract cosmological information~\citep{hamaus_2014_profile,hamaus_2017,hamaus_2020,hamaus_2022}.
The catalogs of random voids and galaxies is characterized by the same redshift dependence of the distributions of voids and galaxies, respectively, as measured in the simulated lightcone, except for the normalization, given by the total number of objects. In particular, we prepared random void catalogs by considering a number of random voids much larger than the corresponding number of voids measured in the simulated lightcone. Moreover, the random catalogs share the same angular footprint as the corresponding void and galaxy catalogs from the lightcone. For the catalog of random voids we also applied to each void center an effective radius, randomly taken from the radius distribution of voids in the lightcone.

The covariance matrix:
\begin{equation}
\label{eq: covariance}
  \mathbf{C}_{ij}=\big\langle \big( \xi^\mathrm{s}(\mathbf{s}_i)-\langle\xi^s(\mathbf{s}_i)\rangle \big) \big( \xi^\mathrm{s}(\mathbf{s}_j)-\langle\xi^\mathrm{s}(\mathbf{s}_j)\rangle \big)  \big\rangle,
\end{equation}
quantifies the uncertainty of the measured VGCF $\xi^\mathrm{s}(s_\perp,s_\parallel)$. In this equation, angled brackets denote the ensemble average of measurements. Since we lack a large enough number of mock catalogs for a precise estimate of the numerical covariance, we estimate the covariance matrix, $\textbf{C}_{ij}$, via the jackknife technique. This methodology relies on ergodicity, which allows us to average measurements across different spatial patches to estimate the covariance matrix. The jackknife method is implemented by measuring $\xi^\mathrm{s}$ through Eq. \eqref{eq: ccf LS} several times, by excluding one (non-overlapping) void at a time. In this way we obtain $N_\mathrm{v}$ samples, where $N_\mathrm{v}$ is the total number of voids, to estimate the corresponding covariance $\textbf{C}_{ij}$.  This methodology has been widely used in prior analyses)~\citep[][]{paz_2013, cai_2016, correa_2019, hamaus_2020, hamaus_2022}, and is suitable here, given the forecasting goal of our paper. We note that previous work showed that the Hartlap correction for systematic biases due to the finite number of independent samples
\citep{hartlap_2007} is expected to be negligible \citep[of a few percent level, see][]{hamaus_2022}, given the number of subsamples, $N_{\rm v}$. It has been demonstrated that, in the limit of large sample sizes, the jackknife technique produces consistent covariance estimates compared to those derived from numerous independent mock catalogs \citep{favole_2021}. Any residual discrepancies between the two methods suggest a slight overestimation of covariances by the jackknife approach, thereby rendering our error forecast conservative.

From the data vector of Eq. \eqref{eq: ccf LS}, and the theoretical model from Eqs. \eqref{eq: xi model M Q} and \eqref{eq: mapping r s M AP}, we can express the corresponding Gaussian likelihood $L(\hat{\xi}^s|\pmb{\Theta})$ of the data $\hat{\xi}^s$ considering the model parameter vector $\pmb{\Theta}=(\beta,\epsilon,\mathcal{M},\mathcal{Q})$ as: 
\begin{align}
     \label{eq: likelihood}
     &\ln{L(\hat{\xi}^s|\pmb{\Theta})}=\\&-\frac{1}{2}\sum_{i,j} \left(\hat{\xi}^s(\mathbf{s}_i)-\xi^s(\mathbf{s}_i|\pmb{\Theta}) \right)\pmb{C}_{ij}^{-1}\left(\hat{\xi}^s(\mathbf{s}_j)-\xi^s(\mathbf{s}_j|\pmb{\Theta}) \right). \nonumber
\end{align}
We evaluate the posterior probability distribution by running MCMC analyses with publicly available {\tt emcee} \citep{emcee} Python package.
We assess the quality of the maximum-likelihood model (best ﬁt) relying on the evaluation of the reduced $\chi^2$ statistic:
\begin{equation}
    \label{eq: chi square}
    \chi^2=-\frac{2}{N_\mathrm{dof}}\ln{L(\hat{\xi}^s|\pmb{\Theta})},
\end{equation}
where the number of degrees of freedom is $N_\mathrm{dof}=N_\mathrm{data}-N_\mathrm{par}$, with $N_\mathrm{data}$ the number of bins and $N_\mathrm{par}$ the number of free parameters.

\begin{figure*}[t!]
\centering\includegraphics[width=\linewidth]{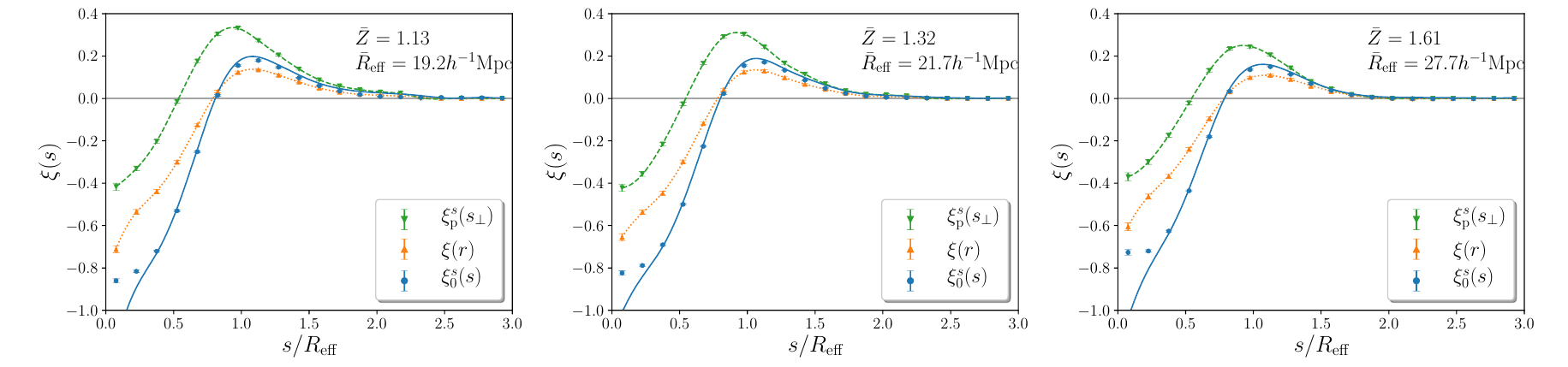}
\caption{Projected void-galaxy cross-correlation function $\xi_p^\mathrm{s}(s_\perp)$ in redshift space (green wedges, interpolated with dashed line) and its real-space
counterpart $\xi(r)$ in 3D after deprojection (orange triangles interpolated with dotted line). The redshift-space monopole $\xi_0^\mathrm{s}(s)$ (blue dots) and its best-fit model based on Eqs. \eqref{eq: xi model M Q} and \eqref{eq: mapping r s M AP} are shown for comparison (solid line). Adjacent bins in redshift increase from left to right, with mean void redshift, $\bar{Z}$, and effective radius, $\bar{R}_\mathrm{eff}$, as indicated in each panel.}
\label{fig: xi_p}
\end{figure*}

In each redshift bin, the data array $\xi^s(s_\perp,s_\parallel)$ is measured in bins of $s_\perp$ and $s_\perp$. We consider 18 linearly equi-spaced bins in each of the two dimensions, resulting in $N_\mathrm{bin}=18 \times 18=324$ bins. It follows that $N_\mathrm{dof}=320$ in each of the considered redshift bins, as $N_\mathrm{par}=4$.
On the one hand this number is much lower than the number of voids per redshift bin considered, therefore guaranteeing enough statistics for a robust estimate of the corresponding covariance matrix, Eq.~\eqref{eq: covariance}; on the other hand, $N_\mathrm{bin}$ is large enough to resolve the features of the 2D VGCF.

\subsection{Results}
\label{subsec: results}

\begin{figure*}[!htbp]
    \centering
    \includegraphics[width=0.95\linewidth]{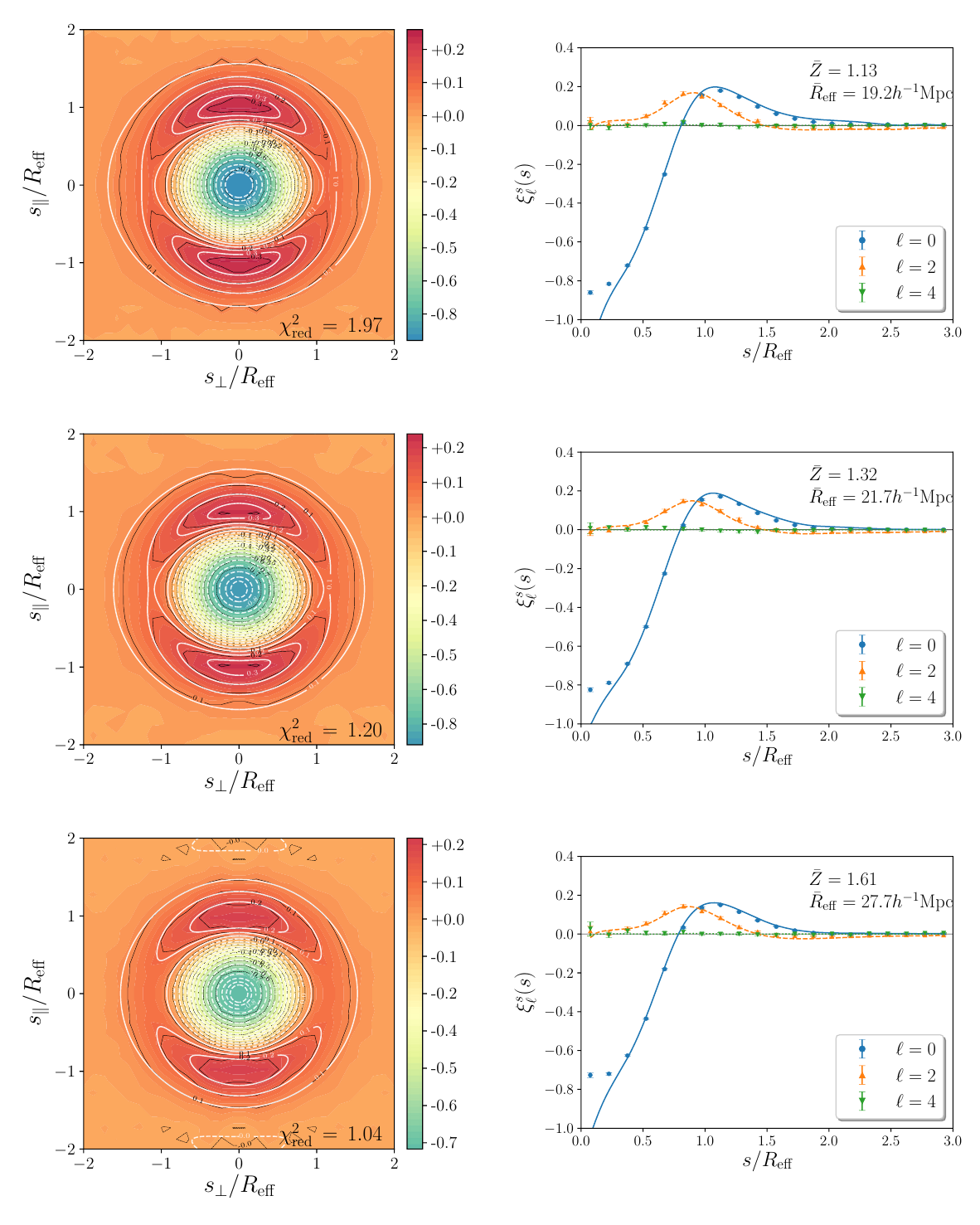}
    \caption{Void-galaxy cross-correlation function in redshift space. Left: $\xi^\mathrm{s}(s_\perp,s_\parallel)$ in 2D (color scale with black contours) and best-fit model from Eqs. \eqref{eq: xi model M Q} and \eqref{eq: mapping r s M AP} (in white contours). Right: monopole (blue dots), quadrupole (orange triangles) and hexadecapole (green wedges) of $\xi^\mathrm{s}(s_\perp,s_\parallel)$ with their best-fit model (solid, dashed, dotted lines). The mean void redshift, $\bar{Z}$, and effective radius, $\bar{R}_\mathrm{eff}$, of each redshift bin are indicated.}
    \label{fig: xi_2d}
\end{figure*}

The VGCF analysis was conducted using the void sample extracted with \vide, a total of $N_\mathrm{v}=82551$ voids, divided into three redshift bins of equal number of voids each, as explained in Section \ref{sec:simu_vide}. Specifically, for the study of the VGCF, the voids underwent post-processing, where a purity cut was applied to avoid the inclusion of spurious Poissonian voids. The cut follows:
\begin{equation}
    R_\mathrm{eff}>N_\mathrm{s}\left( \frac{4\pi}{3}n_\mathrm{g}(Z)\right)^{-1/3},
\end{equation}
where $Z$ represents the redshift of void centers, and the parameter $N_\mathrm{s}$ sets the minimum void size in units of the average tracer separation.
A void catalog with a low $N_\mathrm{s}$ value may be prone to stronger spurious void contamination, i.e. voids misidentified due to the sparsity of tracers, Fingers-of-God, or other systematic effects~\citep{neyrinck_2008_zobov,pisani_2015_peculiar_vel,cousinou_2019,correa_2021,correa_2022}. On the other hand, a high $N_\mathrm{s}$ value can drastically reduce the statistical relevance of the void sample. Choosing the optimal $N_\mathrm{s}$ is a trade-off between these two effects.
In this VGCF analysis, we adopt $N_\mathrm{s} = 3$, resulting in a final sample of $N_\mathrm{v} = 67158$ voids with a minimum effective radius of $8.0 \, h^{-1}$Mpc.

To compute the model for the likelihood analysis via Eqs. \eqref{eq: xi model M Q} and \eqref{eq: mapping r s M AP}, it is essential to determine the stacked density profile or VGCF in real-space, $\xi(r)$. This can be achieved using the deprojection technique outlined in Section~\ref{subsec: theory ccf}. 

We use {\tt Voiager} to compute the LOS integration of $\hat{\xi}^\mathrm{s}(s_\perp,s_\parallel)$ obtaining $\xi^\mathrm{s}_p(s_\perp)$. From this quantity we then obtain $\xi(r)$ via the deprojection technique, Eq. \eqref{eq: xi deprojection}. Integration are performed by interpolating both $\xi^\mathrm{s}_p(s_\perp)$ and $\xi(r)$ with a cubic spline.
Figure \ref{fig: xi_p} illustrates the results for $\xi(r)$ (orange triangles interpolated with dotted line) for each of the three redshift bins, together with the projected void-galaxy cross-correlation function $\xi_p^\mathrm{s}(s_\perp)$ in redshift space (green wedges, interpolated with dashed line) and the redshift-space monopole $\xi_0^\mathrm{s}(s)$ (blue dots) with its best-fit model (blue solid line) based on Eqs.~\eqref{eq: xi model M Q} and \eqref{eq: mapping r s M AP}, which is shown for comparison. 
The statistical noise that could be introduced by the deprojection technique is minimal, due to the large number of voids. However, some residual noise is observed in the innermost bins, where separations from the void center are small and tracers are also sparser,  which affects the accuracy of the deprojection and the subsequent spline interpolation \citep{pisani_2014, hamaus_2020}. Consequently, we omit the first radial bin from our model fits.
To calculate the model for $\xi(r)$ with the deprojection technique, we rely on the data, which introduces its own covariance, leading to a correlation with $\hat{\xi}^\mathrm{s}(s)$. 
Nevertheless, the model considered, Eq.~\eqref{eq: xi model M Q}, considers the amplitude of the VGCF as a free parameter, $\mathcal{M}$. The results are therefore conservative, since any correlation between data and model would reduce the total covariance in our likelihood estimation. 
Besides the deprojection technique presented here, other methods can be used to obtain the $\xi(r)$ in real space. Other studies employ a numerical model based on the measurements of the monopole obtained from simulated data in real space, where available \citep{nadathur_2020, radinovic_2023}. Alternatively, theoretical models could also be used \citep{verza_2024}. By using theoretical models, we could eliminate the potential dependence on the cosmology of the mock data, which would otherwise be introduced when relying on mocks to compute the real-space model. On one hand, the use of a full theoretical model can increase the constraining power; on the other hand the model-free methodology explored in this analysis ensures its robustness, as it is less affected by biases related to unknown systematic effects or potential dependencies on the cosmology of the mock data.

\begin{figure*}[th!]
\centering
\includegraphics[width=\linewidth]{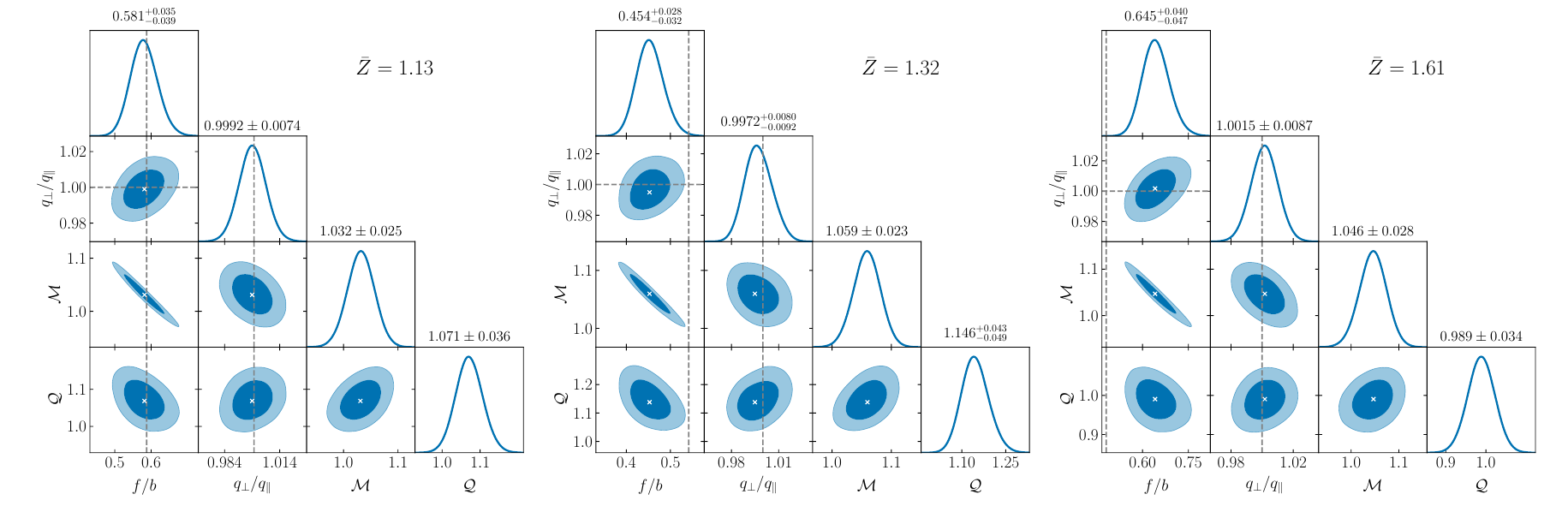}
\caption{VGCF: posterior probability distribution of the model parameters that enter in Eqs. \eqref{eq: xi model M Q} and \eqref{eq: mapping r s M AP}, obtained via MCMC from the data shown on
the left of Figure \ref{fig: xi_2d}. Dark and light-shaded areas represent the $68\%$ and $95\%$ CL, with a cross marking the best fit, dashed lines indicate fiducial values of the RSD and AP parameters. The top of each column states the mean and standard deviation of the 1D marginal distributions. Adjacent bins in void redshift with mean value $\bar{Z}$ increase from left to right, as indicated.}
\label{fig: triangle}
\end{figure*}

\begin{table*}[th!]
    \centering
    \begin{tabular}{
cccccc}\toprule
 $\bar{Z}$ &$b$& $\epsilon \pm \sigma_\epsilon$& $\beta \pm \sigma_\beta$&$f\sigma_8 \pm \sigma_{f\sigma_8}$&$D_\mathrm{A}H/c \pm \sigma_{D_\mathrm{A}H/c}$\\\bottomrule
 $1.13$ &$1.518$& $0.999 \pm 0.007$& $0.581\pm 0.037$&$0.413\pm 0.026$&$1.607 \pm 0.012$ \\\hline
 
        $1.32$ &$1.687$& $0.997 \pm 0.009$ & $0.454 \pm 0.031$ &$0.333 \pm 0.022$&$1.980 \pm 0.017$\\\hline
 $1.61$ &$1.945$& $1.001\pm 0.009$& $0.645\pm 0.044$ & $0.489 \pm 0.033$&$2.622 \pm 0.023$\\\bottomrule
    \end{tabular}
    \caption{VGCF: forecasted constraints on RSD and AP parameters $\epsilon$, $\beta$, $f\sigma_8$, and $D_\mathrm{A}H/c$ (mean values with $68\%$ CL). Results are given in three redshift bins with mean $\bar{Z}$, and large-scale galaxy bias $b$.}
    \label{tab: results_APRSD}
\end{table*}

The data vector in redshift space is constructed using the 2D VGCF, $\xi^\mathrm{s}(s_\perp,s_\parallel)$, estimated with the Landy-Szalay estimator. The measurements are performed across three different redshift bins, as presented in Section \ref{sec:simu_vide}. The resulting VGCF, along with the corresponding multipoles (dots, triangles, and wedges), are shown in Figure \ref{fig: xi_2d}.
We conducted a full MCMC analysis to fit the model described by Eqs.   \eqref{eq: xi model M Q} and \eqref{eq: mapping r s M AP} obtaining the posterior distribution for the model parameters $\pmb{\Theta}=[f/b,\epsilon,\mathcal{M},\mathcal{Q}]$ as illustrated in Figure \ref{fig: triangle}. 
For each MCMC analysis we explore 16 walkers. We then post-process the chains by removing the first 10\% of the steps. To ensure that different steps are uncorrelated, we consider one chain-step every $\tau_{\rm max}/2$, where $\tau_{\rm max}$ is the maximum value of the multi-dimensional time-covariance of the chain, computed with the \citet{goodman_weare_2010} estimator. The resulting length of the chains in the three redshift bins is $\sim 4\times 10^5$. This resulted in a reduced $\chi^2=1.97,\ 1.20,\ 1.04$ in the three redshift bins, respectively. 

We note that the VGCF is sensitive to the growth rate of cosmological perturbations via $f/b$ and to the expansion history of the universe via $\epsilon$. This observable is also sensitive to other cosmological parameters, such as $\sigma_8$ and $h$. However, the ability to constrain these other quantities depends on the specific VGCF model. In this work we considered a model-independent approach, which exclusively probes a subset of background parameters in a robust and unbiased way, through the AP test.

The best-fit models are shown as white contours in the 2D VGCF plots in Figure \ref{fig: xi_2d}. For reference, the corresponding multipole models are also plotted in the panels on the right side. The best-fit models appear to match the data well down to small scales. 
Our results exhibit correlations consistent with those found by \citet{hamaus_2020} and \citet{hamaus_2022}.
Specifically, we observe a weak correlation between $\epsilon$ and $\beta$, and a strong correlation between $\beta$ and $\mathcal{M}$. 
The estimated true values, represented by dashed lines in Figure \ref{fig: triangle}, are computed as follows. 
To compute the true value of $\beta$, both the values for $f(z)$ and $b(z)$ are required. The growth rate $f$ is computed via the following relation, with cosmological parameters provided by the simulation: 
\begin{equation}
    \label{eq: f(z)}
    f(z)\simeq \left[ \frac{\Omega_\mathrm{m}(1+z)^3}{H^2/H_0^2}\right]^\gamma,
\end{equation}
with a growth index $\gamma\simeq 0.55$ \citep{lahav_1991, linder_2005}.
For the bias $b(z)$ we assume the relation in \citet{wang_2022}. The parameter $\epsilon$ is expected to be equal to 1, as we used the cosmology of the simulation to convert angles and redshifts into distances, introducing no AP effect. The parameters $\mathcal{M}$ and $\mathcal{Q}$ do not have specific values, and their distribution is not relevant to the cosmological interpretation of the posteriors (given their use to account for potential biases due to e.g., spurious voids contamination and noise in the deprojection).

The relative precision of both $\beta$ and $\epsilon$ varies across the redshift bins, with values ranging from $6.4\%$ to $6.8\%$ for $\beta$ and from $0.7\%$ to $0.9\%$ for $\epsilon$. Notably, $\epsilon$ is measured with high precision and accuracy in all redshift bins, with deviations of only $0.1\sigma_\epsilon$ to $0.3\sigma_\epsilon$ from its true value.
We note that for $f/b$ some discrepancies may arise---observable here thanks to the high statistical power of the Roman void sample leading to tight error bars--- and that could lead to biases in the inferred $f \sigma_8$ due to two reasons. First, the bias parameter can be different from its large-scale counterpart (obtained with galaxies two-point correlation function or power spectrum computations, see Section~\ref{subsec: theory ccf}), especially in small voids \citep{pollina_2017, verza_2022}. In this case the linear bias approximation could still be valid, but the linear relation may be verified with a different slope. Second, projection effects in the parameter space, due to the strong degeneracy between $f/b$ and ${\cal M}$, can affect the inferred value of $f \sigma_8$.
It is important to note that these deviations in $\beta$ do not significantly affect the overall quality of the fit. This is primarily due to the role played by the nuisance parameters $\mathcal{M}$ and $\mathcal{Q}$, which help to mitigate the impact of the deviations in $\beta$ without substantially influencing the precision or accuracy of $\epsilon$.

From the posteriors of $\beta \equiv f/b$ and $\epsilon\equiv q_\perp/q_\parallel$, it is possible to derive constraints on $f\sigma_8$ and $D_\mathrm{A} H$. For the former, we assume that $\xi(r)$ is proportional to $b\sigma_8$, and thus we multiply $f/b$ by the underlying value of $b\sigma_8$ provided by the mock. Additionally, we assume that the relative precision on $f/b$ and $f\sigma_8$ is the same. 
Furthermore, we neglect the dependence on $h$, which enters into the definition of $\sigma_8$ and should ideally be marginalized over \citep{sanchez_2020}. For the latter case, we compute $D_\mathrm{A} H$ by multiplying $\epsilon$ by the fiducial $D_\mathrm{A}H$, following Eq. \eqref{eq: epsilon}.  We summarize all the results in Table \ref{tab: results_APRSD}.
As expected, the discrepancies observed in the values of $\beta$ in the second and third redshift bins propagate to the estimated values of $f\sigma_8$.

The measurements of $f\sigma_8$ and $D_A H$ as a function of redshift can be used to constrain cosmological models, for example, by inverting Eqs. \eqref{eq: DA(z)} and \eqref{eq: H(z)}. As in the analysis of the VSF, we explore three different cosmological models. 
The first model is the $\Lambda$CDM model, where the parameter space is $\pmb{\Theta} = [\Omega_\mathrm{m}]$, probed by the AP test. We use flat priors for $\Omega_\mathrm{m}$ in the range $(0.15, 0.65)$. The second model is the $w$CDM model, with parameters $\pmb{\Theta} = [\Omega_\mathrm{m}, w]$, combining the $\Lambda$CDM priors with a flat prior on $w$ in the range $(-3, 2)$. The third model is the $w_0w_\mathrm{a}$CDM model, with parameters $\pmb{\Theta} = [\Omega_\mathrm{m}, w_0, w_\mathrm{a}]$, and flat priors identical to those of $w$CDM, along with $w_\mathrm{a}$ in the range $(-8, 5)$.

\begin{table}[t!]
    \centering
    \begin{tabular}{cccc}\toprule
 Model&$\Omega_\mathrm{m}$& $w_0$& $w_\mathrm{a}$\\\hline\hline
 $\Lambda$CDM&$0.308^{+0.007}_{-0.009}$& $-1$& $0$\\\hline
 
        $w$CDM&$0.303^{+0.013}_{-0.015}$& $-1.05^{+0.25}_{-0.27}$& $0$\\\hline
 $w_0w_\mathrm{a}$CDM&$0.295^{+0.215}_{-0.070}$& $-1.06^{+1.80}_{-0.44}$& $1.04^{+0.70}_{-6.64}$\\\bottomrule
    \end{tabular}
    \caption{VGCF: forecasted constraints on cosmological parameters, $\Omega_\mathrm{m}$, $w_0$, and, $w_\mathrm{a}$, estimated assuming three different cosmological models: $\Lambda$CDM, $w$CDM,  and $w_0w_\mathrm{a}$CDM.}
    \label{tab: results_cosmo}
\end{table}

\begin{figure}
    \centering
    \includegraphics[width=1\linewidth]{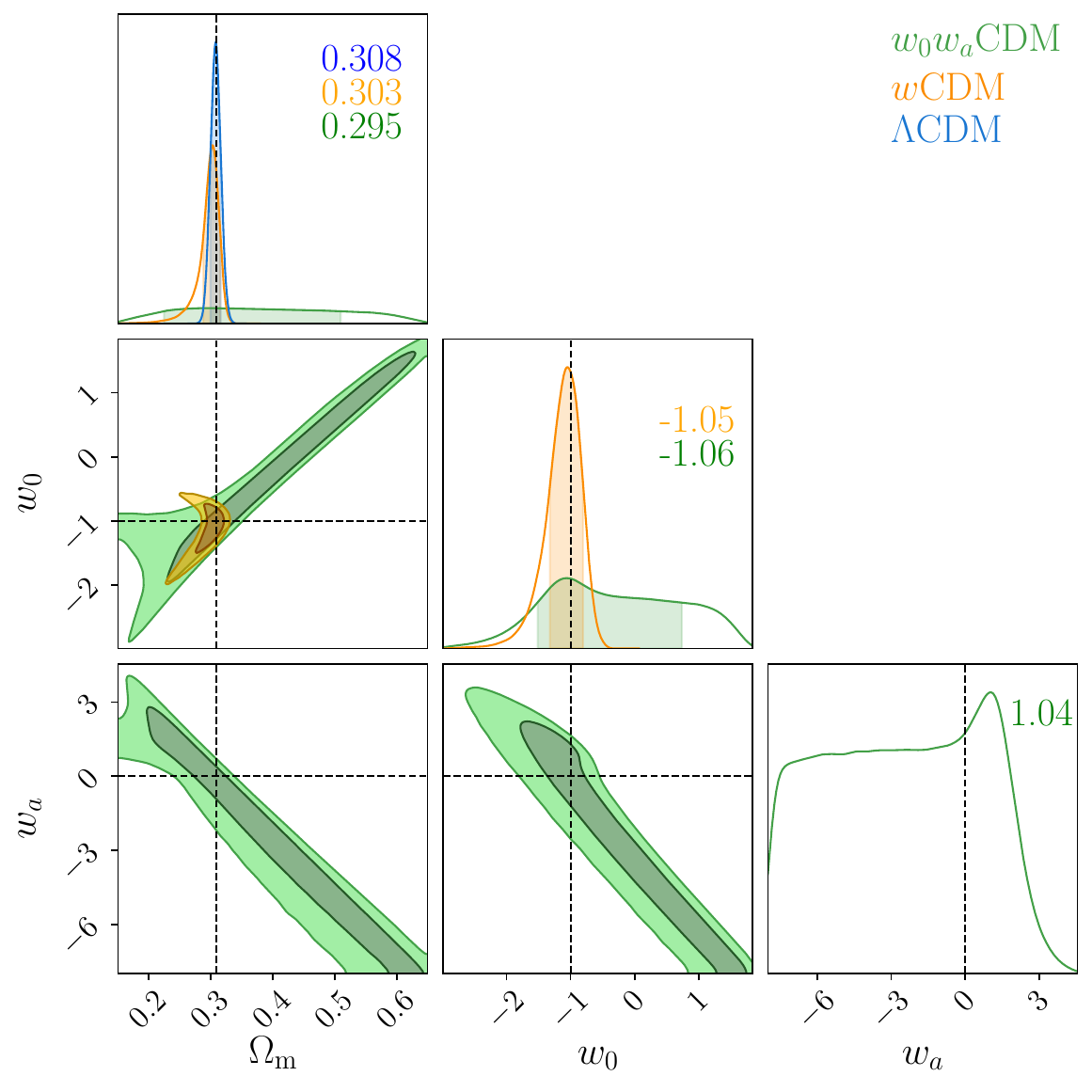}
    \caption{VGCF: expected constraints on $\Omega_\mathrm{m}$, $w_0$, and $w_\mathrm{a}$ for the three different cosmologies tested here. In blue we show the $\Lambda$CDM model, while orange and green correspond to $w$CDM and $w_0w_\mathrm{a}$CDM respectively. Dashed lines indicate the true values. 
    Numbers in color correspond to the estimated values for the parameters, for the different cosmological models.}
    \label{fig: triangle_cosmology}
\end{figure}

Figure \ref{fig: triangle_cosmology} shows the 2D  marginalized posterior distributions of the cosmological models explored: $\Lambda$CDM (blue), $w$CDM (orange), and $w_0w_\mathrm{a}$CDM (green). The shaded regions represent the $68\%$ confidence level, while the outer contours indicate the $95\%$ confidence level. The true values of the parameters are marked by gray dashed lines.
Table \ref{tab: results_cosmo} lists the best-fit values and the 1D $68\%$ confidence intervals for each cosmological parameter in each model.
In a $\Lambda$CDM model, the one assumed for the simulation, we recover the value of $\Omega_\mathrm{m}$ with a precision of $\simeq 3\%$ that lays in a $0.1\sigma$ interval from the true value.
In the second scenario, a $w$CDM model, $\Omega_\mathrm{m}$ is estimated with a precision of $\simeq 6.6 \%$ and its distance from the true value is $0.3\sigma$. 
The third and final scenario, the $w_0w_\mathrm{a}$CDM model, is more complex, leading to larger uncertainties and making the best-fit values more challenging to estimate from the posteriors. Typically in future analyses a wider redshift range, more redshift bins, and/or probe combination can break degeneracies. Moreover, in future studies, we should account for projection effects in the posterior distribution \citep{raveri_2024}.

\begin{table*}
\centering
\begin{tabular}{ccccccc}
\toprule
\noalign{\vspace{0.075cm}}
Model & & $\Omega_{\rm m}$ & $\sigma_8$ & $h$ & $w_0$ & $w_{\rm a}$  \\ 
\noalign{\vspace{0.1cm}}
\toprule
\noalign{\vspace{0.15cm}}
True & & $0.3089$ & $0.8147$ & $0.6774$ & $-1$ & $0$ \\ 
\noalign{\vspace{0.4em}}
\cline{1-7}
\noalign{\vspace{0.4em}}
\multirow{3}{*}{$\Lambda$CDM} & opt. & $0.3086^{+0.0077}_{-0.0080}$ & $0.8145^{+0.0258}_{-0.0272}$ & $0.6764^{+0.0164}_{-0.0168}$ & $-1$ & $0$  \\ 
\noalign{\vspace{0.4em}}
\cline{3-7}
\noalign{\vspace{0.4em}}
& pess. & $0.3100^{+0.0066}_{-0.0093}$ & $0.8212^{+0.0211}_{-0.0336}$ & $0.6726^{+0.0212}_{-0.0147}$ & $-1$ & $0$ \\ 
\noalign{\vspace{0.4em}}
\cline{1-7}
\noalign{\vspace{0.4em}}
\multirow{3}{*}{$w$CDM} & opt. & $0.3095^{+0.0052}_{-0.0124}$ & $0.8167^{+0.0186}_{-0.0392}$ & $0.6757^{+0.0215}_{-0.0144}$ & $-1.023^{+0.171}_{-0.113}$ & $0$ \\ 
\noalign{\vspace{0.4em}}
\cline{3-7}
\noalign{\vspace{0.4em}}
& pess. & $0.3089^{+0.0063}_{-0.0121}$ & $0.8149^{+0.0226}_{-0.0389}$ & $0.6801^{+0.0189}_{-0.0208}$ & $-1.041^{+0.195}_{-0.101}$ & $0$ \\ 
\noalign{\vspace{0.4em}} 
\cline{1-7}
\noalign{\vspace{0.4em}}
\multirow{3}{*}{$w_0 w_{\rm a}$CDM} & opt. & $0.3238^{+0.0250}_{-0.0262}$ & $0.8601^{+0.0721}_{-0.0864}$ & $0.6584^{+0.0246}_{-0.0297}$ & $-0.934^{+0.216}_{-0.182}$ & $-0.47^{+0.93}_{-0.93}$ \\ 
\noalign{\vspace{0.4em}}
\cline{3-7}
\noalign{\vspace{0.4em}}
& pess. & $0.3313^{+0.0247}_{-0.0310}$ & $0.8760^{+0.0780}_{-0.0990}$ & $0.6563^{+0.0337}_{-0.0370}$ & $-0.904^{+0.240}_{-0.208}$ & $-0.97^{+1.38}_{-1.23}$ \\  
\noalign{\vspace{0.15cm}}
\hline
\bottomrule
\end{tabular}
\caption{Combination of the VGCF and VSF: maximum posterior distribution values and 1D 68\% CL interval of the cosmological parameters explored for each considered cosmological model (we show both the optimistic and pessimistic scenarios).}
\label{tab:combo_constraints}
\end{table*}

To conclude this Section, it is worth mentioning that we have relied on the method presented by \cite{hamaus_2020}, that significantly improves the constraining power of the VGCF. For a more conservative forecast we have allowed the parameters $\mathcal{M}$ and $\mathcal{Q}$ to vary. While leaving those parameters free to vary is a conservative option, we notice here that the full power of the method can be reached by fixing those parameters, that is when a better understanding of spurious voids and other aforementioned effects is obtained. Further tests to increase our understanding of such parameters, which we aim to conduct in future studies, include the study of the dependence of $\mathcal{M}$ and $\mathcal{Q}$ on the cosmological and structure formation models assumed in the mocks (e.g., on the modelling of galaxy properties). Alternatively, for tighter constraints it is possible to rely on a set of mocks to calibrate the values of $\mathcal{M}$ and $\mathcal{Q}$ that can then be held fixed during the analysis, at the expense of trusting the features of the mocks used for calibration. Given these points, until a better understanding is reached, calibrated results should be considered less robust; therefore, we do not explore the possibility of calibrating them here.

\section{Discussion and conclusion}\label{sec:conclu}

\begin{figure*}[th!]
\centering
\includegraphics[width=\linewidth]{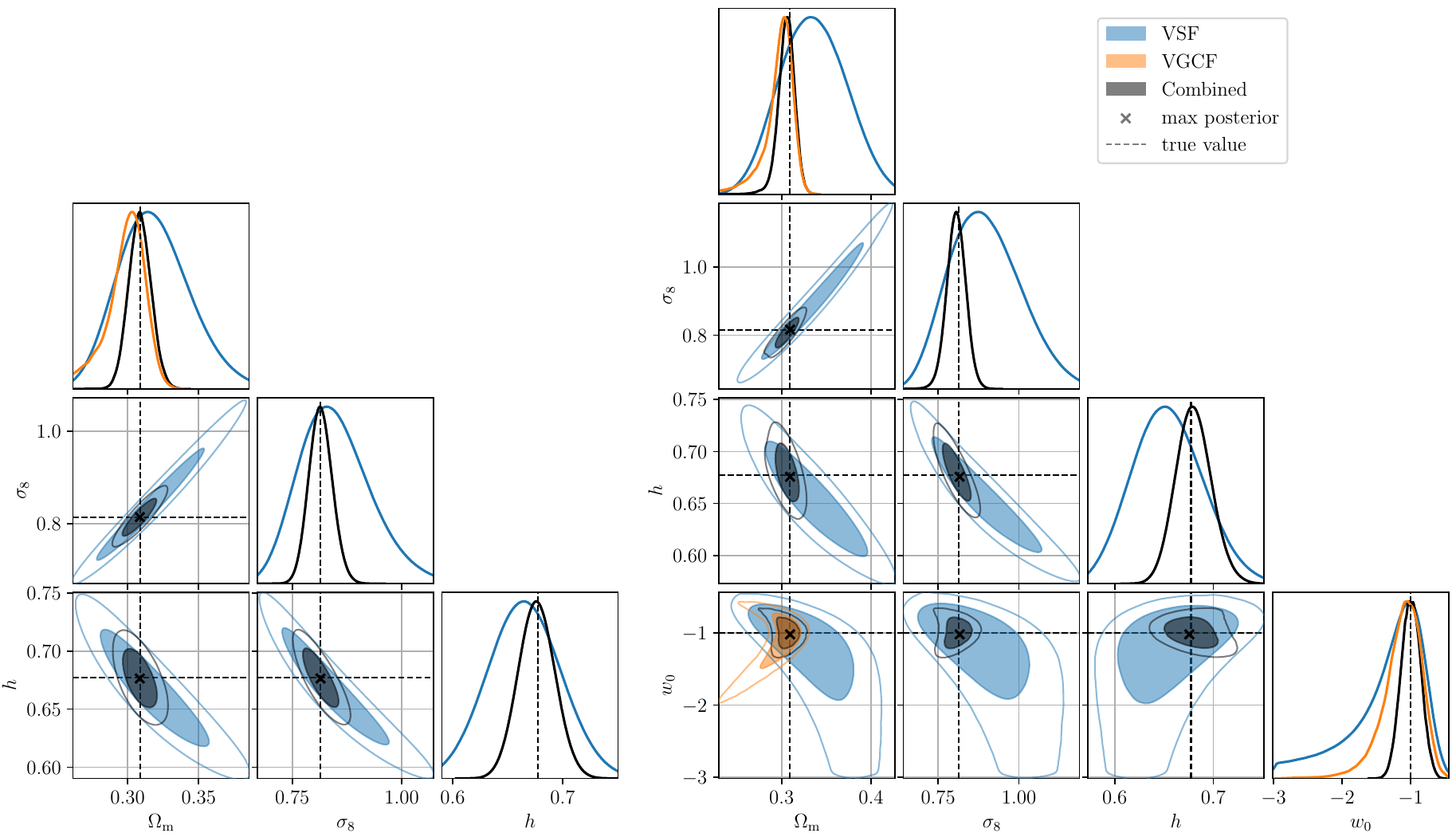}
\caption{Combination of the VGCF and VSF: posterior distributions of the cosmological parameters explored in the $\Lambda$CDM (left) and the $w$CDM model (right), obtained from the VSF (blue), VGCF (orange), and combining the two statistics (dark gray). The filled internal region shows the 68\% CL, the outer line shows the 95\% CL. Black dashed lines indicate the true values, crosses correspond to the maximum of the posterior distributions.}
\label{fig:combo_LCDM}
\end{figure*}  

\begin{figure*}
\centering
\includegraphics[width=0.85\linewidth]{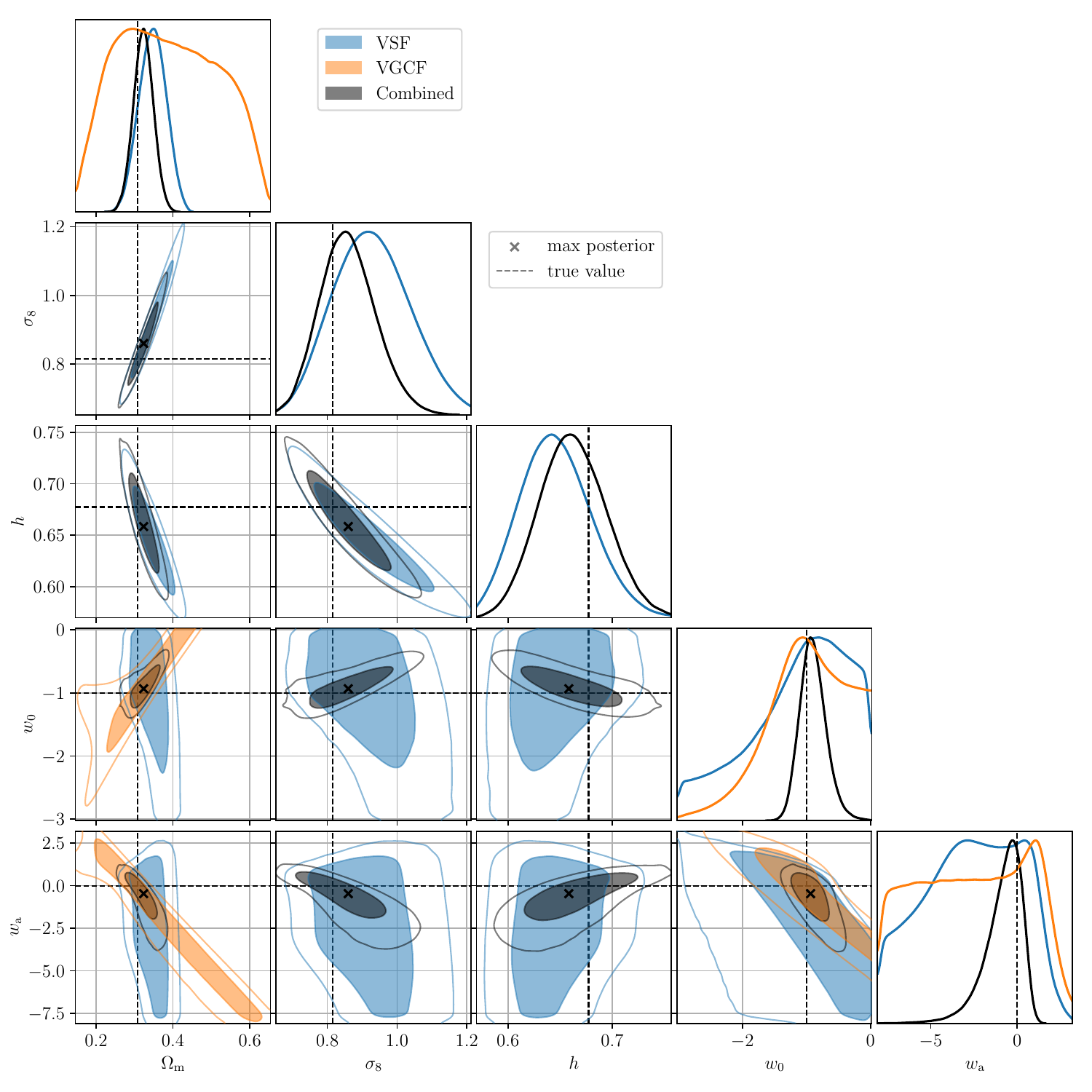}
\caption{Combination of the VGCF and VSF: posterior distributions of the cosmological parameters explored in the $w_0w_{\rm a}$CDM model, obtained from the VSF (blue), VGCF (orange), and combining the two statistics (dark gray). The plot is organized as Fig~\ref{fig:combo_LCDM}.}
\label{fig:combo_w0waCDM}
\end{figure*}

In this work we present a comprehensive cosmological forecast from void statistics to be measured in the Roman reference HLSS survey. We build void catalogs relying on a Roman-like mock and employ state of the art models to extract cosmological constraints from the VSF and VGCF. Our results showcase an impressive constraining power from both statistics.  
The full power of the different void statistics can be confirmed by considering voids' probe combination, as the two statistics explored here respond differently to cosmology.
The VSF is sensitive to both the cosmological fluctuations, through the linear matter power-spectrum, and the background expansion of the Universe (see Sections~\ref{subsec:vsf_model}--\ref{subsec:vsf_analysis}). The VGCF is a sensitive probe of the expansion history of the universe, through the AP test. 
Previous work suggests that the covariance between the VSF and the VGCF is low~\citep{kreisch_2022, contarini_2024}, therefore the constraining power of the joint analysis can be safely estimated by combining the two independent posterior distributions.
To showcase the constraining power of a joint void-analysis, Figures~\ref{fig:combo_LCDM} 
and \ref{fig:combo_w0waCDM} represent the overlapping forecast contours for the cosmological models and parameters explored in this work, considering the optimistic case for the VSF. We list in Table 
\ref{tab:combo_constraints} the best-fit value and 1D 68\% CL for each cosmological parameter from the joint distribution, for both the optimistic and pessimistic scenario. The left panel of Figure~\ref{fig:combo_LCDM} shows the $\Lambda$CDM case, with the marginalized 2D posterior distribution of the VSF in blue, the VGCF in orange, and the combined posterior distribution in dark gray. It should be noted that the narrower posterior distribution for $\Omega_{\rm m}$ in the VGCF case leads to a tightening of constraints even for parameters that are not probed by the VGCF, i.e. $\sigma_8$ and $h$ (or $H_0$). The same observation is valid also for the $w$CDM case, Figure~\ref{fig:combo_LCDM} left panel. Moreover, it is important to notice the different orientation, i.e. the complementarity, in the $w-\Omega_{\rm m}$ plane. For the $w_0w_{\rm a}$CDM case, shown in Figure \ref{fig:combo_w0waCDM}, the situation is inverted. The VGCF, probing the background expansion of the Universe, shows wider contours, in particular for $\Omega_{\rm m}$. The corresponding VSF posterior distribution shows a higher constraining power, probably due to the fact that, together with its impact on the expansion history of the Universe, the dark energy equation of state also affects the growth of structure, i.e. $\sigma_8$, to which the VSF is directly sensitive, while the AP test, performed with the VGCF, is not.
Additionally, we wish to notice that, while selecting three bins is an effective choice for the analysis of the VGCF when considering the $w$CDM case, it may be harder to break degeneracies with three bins when considering the $w_0w_{\rm a}$CDM case. Nevertheless, the combination of the VSF and the VGCF shows an interesting complementarity for $\Omega_{\rm m}$ and the dark energy parameters, $w_0$ and $w_{\rm a}$, leading to strong constraints for the dynamical dark energy parameters as well. 

In future works, when combining the various void statistics, we plan to rely on mock catalogs spanning various cosmological realizations and to consider various galaxy-halo connection models, such as SAM, halo-occupation distribution (HOD), and sub-halos matching techniques (SHAM), and spanning the corresponding parameters. This will allow to properly model the covariance, for both the VSF, the VGCF, and the cross-covariance, modeling the cosmological dependence and marginalizing over galaxy properties, represented by the galaxy-halo connection models. 
Crucially, we expect the use of mock galaxy catalogs to allow a better understanding of the free parameters of our models, such as the redshift and tracer dependency for the VSF effective barrier in Eq.~\eqref{eq:moving_b}, and the nuisance parameters ${\cal M}$ and ${\cal Q}$ for the VGCF model in Eq.~\eqref{eq: xi model M Q}. Such understanding, together with the increase in the constraining power for both statistics, will enhance the robustness of our models, alongside with accounting for observational effects (e.g., through the use of realistic survey masks). 
In the current cosmology landscape, if dynamical dark energy is confirmed~\citep{DESI_BAO_2024}, the community will focus on pinning down the dark energy equation of state, and Roman voids are expected to be a powerful, independent probe in this context. 
With a first, comprehensive analysis of a Roman-like mock, this work paves the way to using Roman voids to independently constrain cosmological parameters with tight precision.

\begin{figure*}[th!]
\centering
\includegraphics[width=0.85\linewidth]
{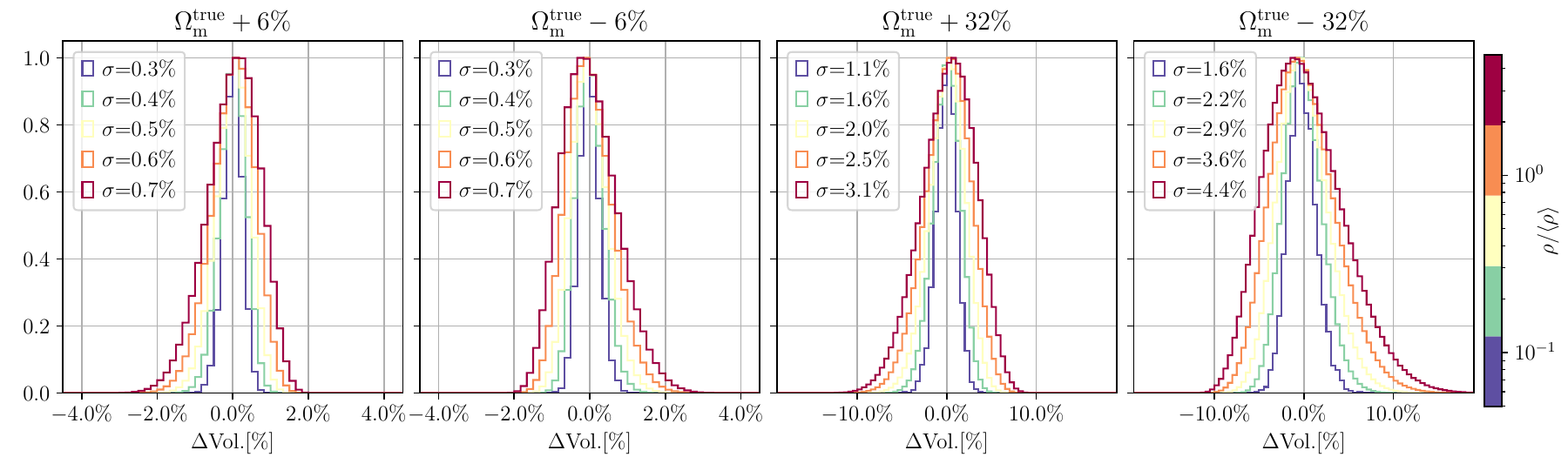}
\caption{Histograms showing the dispersion of the relative difference between the volume of the Voronoi cells in the true cosmology and different fiducial cosmologies corrected with the AP factor. The histograms are normalized with respect to the maximum, and refer to Voronoi cells of galaxies with redshift $z \in [1.53.1.895]$. Each panel shows a different fiducial cosmology. The histograms of different color show the results for Voronoi cells with different normalized density,  $\rho / [\langle \rho \rangle(z)]$. The color, from blue to dark red, refers to the following normalized density binning: (0, 0.049,0.15, 0.49, 1.5, 4.8, 15).}
\label{fig:AP_vornoi}
\end{figure*}

\vspace{4em}
GV acknowledges NASA grant EUCLID12-0004. GV and AP acknowledge support from the Simons Foundation to the Center for Computational Astrophysics at the Flatiron Institute. 
AP acknowledges support from the European Research Council (ERC) under the European Union's Horizon programme (COSMOBEST ERC funded project, grant agreement 101078174), as well as support from the french government under the France 2030 investment plan, as part of the Initiative d'Excellence d'Aix-Marseille Université - A*MIDEX AMX-22-CEI-03. YW's work is funded in part by NASA grant \#80NSSC24M0021. Part of this work was done at Jet Propulsion Laboratory, California Institute of Technology, under a contract with the National Aeronautics and Space Administration (80NM0018D0004).
The authors are grateful to Enzo Branchini, Hélène Courtois, Shirley Ho, Elena Sarpa and David Spergel for useful discussions.
This paper relies on the public code {\tt Voiager}\footnote{\url{hhttps://voiager.readthedocs.io/}}.

\appendix

\section{Impact of the fiducial cosmology on the void finding procedure}\label{appendix}

In this Appendix we deepen the discussion of Section~\ref{subsec:vsf_analysis} about the impact of the assumed fiducial cosmology on the Voronoi tessellation, on the void finding procedure and on the final void catalogs.
From a theoretical perspective, the effect of using a fiducial cosmology different from the true one in producing the void catalogs can be corrected considering the AP transformation, based on Eq.~\eqref{eq:AP_par_perp}. We now show---both analytically and numerically---first, how each Voronoi cell transforms under a change in the assumed cosmology and, second, how the global void catalogs transform, considering both \vide and threshold voids.

For this investigation, we ran \vide and find the corresponding threshold voids on lightcones using various fiducial cosmologies to convert the galaxy coordinates to comoving distances. Instead of using the 2000 square degrees galaxy lightcone~\citep{zhai_2021}, we use for this investigation the halo lightcone of the {\it AbacusSummit} huge simulations~\citep{Garrison_2021,
Maksimova_2021,Hadzhiyska_2022}. These are built using a simulation box of 7600 $h^{-1}$Mpc side, $8640^3$ particles with mass of $2.1 × 5 × 10^{10} h^{-1}M_\odot$, and cosmological parameters $[\Omega_{b}h^2, \Omega_{\rm cdm}h^2, h, 10^9A_{\rm s}, n_{\rm s}, w_0, w_{\rm a}]= [0.02237, 0.12, 0.6736, 2.083, 0.9649,-1, 0]$. For the analysis we select halos with mass $3 \times 10^{12} h^{-1}M_\odot$, for a total number of halos of $\sim 1.8 \times 10^{8}$.
This is a full-sky lightcone with redshift extending up to $z \simeq 2.5$. Therefore, the results presented here are general and the associated uncertainty corresponds to the cosmic variance.

The AP transformation of Eq.~\eqref{eq:AP_par_perp} is a local linear transformation of the coordinate position. In particular, it is defined in the limit $\Delta z \ll 1$, i.e. where the comoving distance can be expanded as
\begin{equation}
D_{\rm A}(z+\Delta z) = \int_0^z \frac{c{\rm d}z'}{H(z')} + \frac{c\Delta z}{H(z)} + {\cal O}\left[(\Delta z)^2\right]\,
\end{equation}
and for $\Delta \theta \ll 1 $, i.e. where the $\arctan (\Delta \theta) = \Delta \theta + {\cal O}[(\Delta \theta)^2]$. From this, Eq.~\eqref{eq:AP_par_perp} immediately follows. Since the typical linear extension of a Voronoi cell is of the order of the mean galaxy separation, which in our case is always less than 10 $h^{-1}$Mpc, the above condition is always satisfied. 

Under a linear transformation, the volume of a polyhedron transforms according to the determinant of the change of coordinates, which, for the AP transformation is $q_\parallel q_\perp^2$, as described in Section~\ref{subsec:vsf_analysis}. 
The transformation of a single Voronoi cell is a little more subtle, as an anisotropic linear transformation may change the angles among the faces of the cell, and consequently the intersection regions. It follows that Voronoi cells detected in the fiducial cosmology and transformed in the true one, can show slightly different angles with respect to the ones directly detected in the true cosmology. We verified on simulations that this effect is subdominant and its impact on the \vide and thresholding  procedure is negligible, for three reasons. First, we numerically check that the change in the cell sizes can be described as noise with amplitude always below 1\% for physically motivated physical scenarios (see below). 
Second, even if the shape of single Voronoi cells slightly changes when computed using different fiducial cosmologies; if we perform the tessellation at a given cosmology and then transform this initial tessellation with AP (instead of computing it again in the other cosmology), it remains an excellent field estimator. 
Therefore the relations described in Section~\ref{subsec:vsf_analysis} remain exact. Third, when identifying voids, several cells are added together, therefore any random noise in the volume of each single Voronoi cell sums to zero.

\begin{figure}[t!]
\centering
\includegraphics[width=\linewidth]{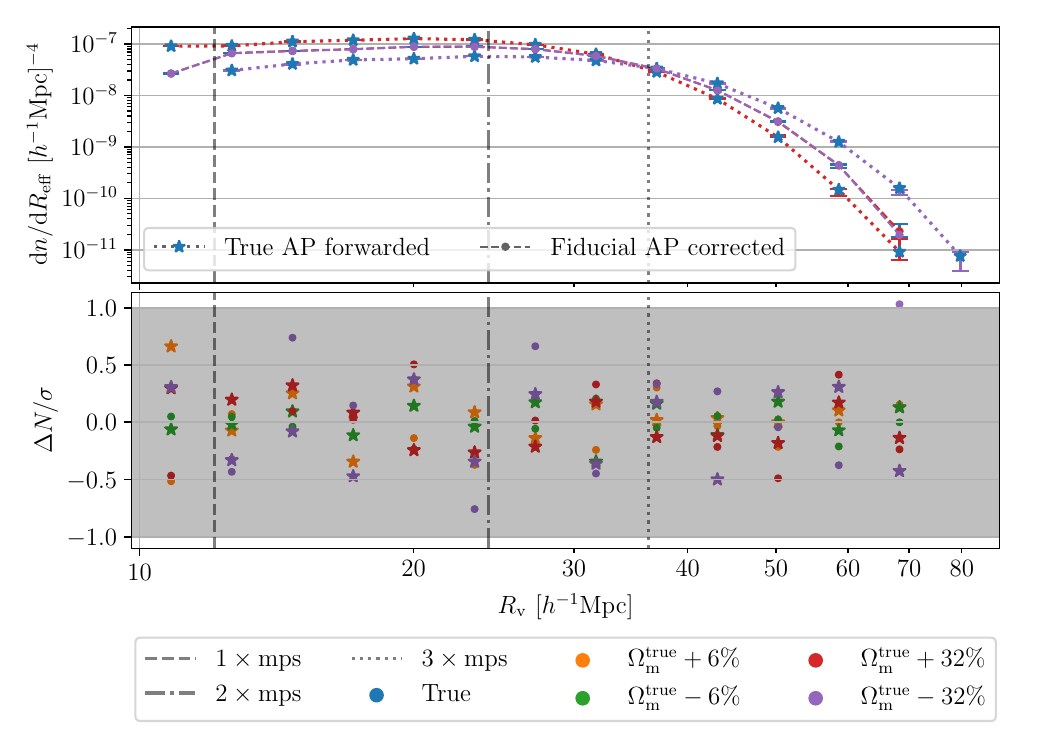}
\caption{VSF of threshold voids with $\delta_{\rm v,g}=-0.7$ detected in the Abacus lightcone in a redshift bin of $z \in [1.54,1.895]$ in various fiducial cosmologies, as listed in the legend. Upper panel: the error bars show the VSF measured assuming various fiducial cosmologies (also listed in the legend), the uncertainty is Poissonian. The colored circles connected with a dashed line of the same color show the VSF measured in fiducial cosmologies, different from the true one, and corrected using AP. The blue stars connected with a blue line show the VSF measured in the true cosmology and forwarded to the fiducial cosmologies. To avoid overpopulating the plot, we only show here the two most extreme cases, $\Omega^{\rm fid.}_{\rm M}=\Omega^{\rm true}_{\rm M}\pm 32\%$.
Lower panel: ratio between the AP corrected VSF in the fiducial cosmologies and the one in the true cosmology (circles); ratio between the VSF in the true cosmology forwarded to the fiducial with AP with respect to the ones measured in the corresponding fiducial cosmology (stars). The ratios are shown in units of the Poissonian uncertainty, which corresponds to the cosmic variance. The vertical dashed, dashed-dotted, and dotted lines correspond to 1, 2, and 3 times the mean galaxy separation, respectively.}
\label{fig:AP_VSF_threshold}
\end{figure}

The comparison between the volume of the Voronoi cells computed in the true cosmology and the one in the fiducial cosmology is obtained by performing the ratio of the volume of the Voronoi cells with the same central galaxy in the true and fiducial cosmologies, and then compared to the theoretical AP factor $q_\parallel q_\perp^2$. Figure~\ref{fig:AP_vornoi} shows the distribution of this dispersion for four fiducial cosmologies different from the true one, in the redshift range of $z \in [1.54,1.895]$. The four considered cosmologies are flat-$\Lambda$CDM, where the $\Omega_{\rm m}$ parameter has been increased or decreased by a factor of 6 and 32\%, represented in each panel.  The choice of the redshift bin is a trade-off between the intensity of the AP effect (that varies with redshift) and the available statistics. We verified that at higher redshift the distributions remain almost unchanged, while the width of the distributions shrinks when the redshift decreases, going to 0 for $z \rightarrow 0$. Different colors, from blue to dark red, represent the distribution of Voronoi cells binned according to their normalized density, $\rho / [\langle \rho \rangle(z)]$, in 5 logarithmically spaced bins: (0, 0.049, 0.15, 0.49, 1.5, 4.8, 15). We note that the width of the distribution increases with the density of the Voronoi cells. For voids, we are interested in the three lowest density bins.

\begin{figure}[t!]
\centering
\includegraphics[width=\linewidth]{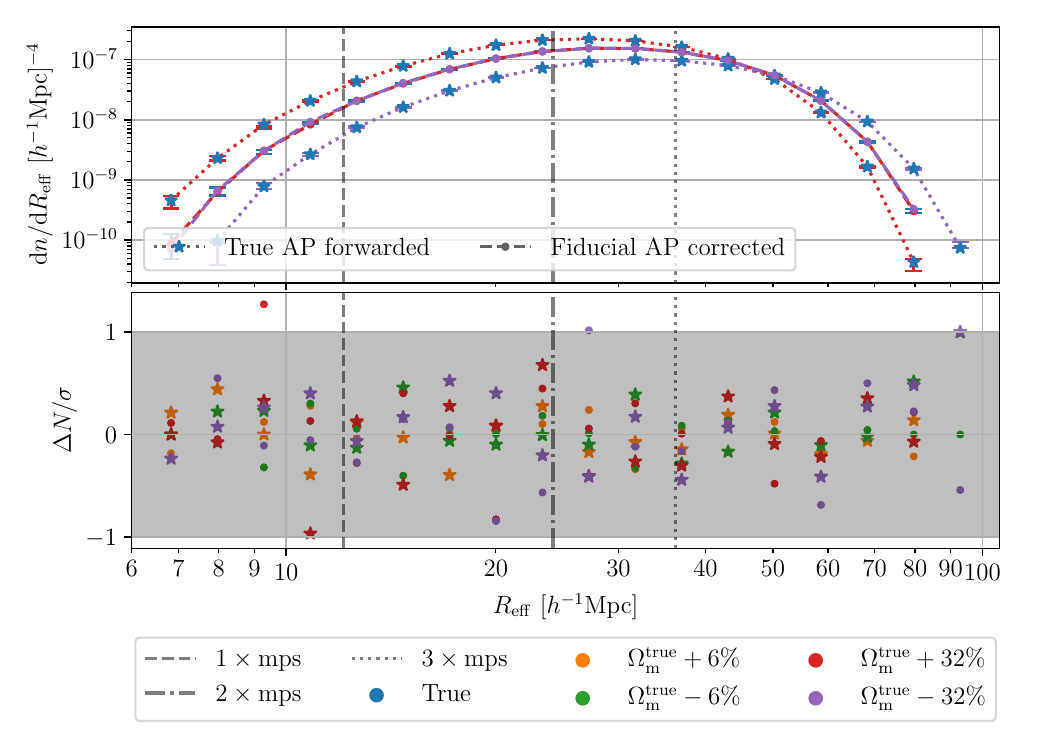}
\caption{VSF of \vide voids detected in the Abacus lightcone in a redshift bin of $z \in [1.54,1.895]$ in various fiducial cosmologies, as listed in the legend. The plot is organized as Figure~\ref{fig:AP_VSF_threshold}}
\label{fig:AP_VSF_vide}
\end{figure}

The understanding of how voids transform under a fiducial cosmology different from the true one follows from the same assumption used for the Voronoi tesselation.
The redshift and angular extension of the considered voids always satisfy the conditions $\Delta z \ll 1$ and $\Delta \theta \ll 1$. As a consequence, the identification of threshold voids is expected to be conserved when the fiducial cosmology changes. In particular, for threshold voids the galaxies belonging to each void are not expected to depend on the assumed fiducial cosmology. This means that the redshift and angular coordinates are expected to be independent of the fiducial cosmology, as long as the AP approximation is valid, while the volume transforms according to AP, as demonstrated in Section~\ref{subsec:vsf_analysis}. Moreover, the small noise that arises when AP-correcting the Voronoi volumes from the fiducial to the true cosmology cancels out by summing the volume of the many cells constituting a void. 

By comparing voids matching the ID of the cell with the lowest density, we explicitly verified that the identification of voids is conserved when changing the fiducial cosmology. Notably, this simple method matches 97\% of the voids detected in each of the different fiducial cosmologies with voids detected in the true cosmology case. 

Figure~\ref{fig:AP_VSF_threshold} shows the result for threshold voids in the {\it AbacusSummit} huge lightcone, with $\delta_{\rm v,g}=-0.7$, obtained following the same procedure described in Section~\ref{subsec:vsf_cat}. As for the single Voronoi cells case, we choose the redshift bin of $z \in [1.54,1.895]$, as a trade-off between the AP effect and the statistics of the sample. Other redshift bins show analogous results. The error bars in the upper panel show the VSF measured assuming 3 fiducial cosmology, the true one and $\Omega^{\rm true}_{\rm m} \pm 32\%$, the error bars are Poissonian and correspond to the cosmic variance. We show the true cosmology and the two most extreme fiducial cosmologies considered to avoid an over-populated plot. 
The colored circles connected with a dotted line of the same color show the VSF measured in fiducial cosmologies, different from the true one, and corrected using the AP. The blue stars connected with a colored dashed line show the VSF measured in the true cosmology and forwarded to the fiducial ones (labeled as the color indicates). The lower panel shows the relative difference between the AP corrected VSF in the fiducial cosmologies and the one in the true cosmology (circles); and between the VSF in the true cosmology forwarded to the fiducial with AP with respect to the ones measured in the corresponding fiducial cosmology (stars). The results are expressed in units of the Poissonian uncertainty. All the fiducial cosmologies considered are shown. It can be seen that the differences among the various corrected or forwarded VSFs using AP are consistent with each other within $1\sigma$, that is, within the cosmic variance. The vertical dashed, dash-dotted, and dotted lines correspond to 1, 2, and 3 times the mean galaxy separation of the redshift bin, respectively.

The same behavior is expected also for watershed voids because a linear transformation of galaxy coordinates does not change the topological structure, on which watershed void finders rely. Since the AP transformation can be considered valid on scales larger than the typical void size, the watershed identification is retained, except from subdominant effects. This is confirmed by numerical analyses on the {\it AbacusSummit} huge lightcone, showed in Figure~\ref{fig:AP_VSF_vide}. This plot shows the effect of various fiducial cosmologies on the VSF of \vide voids, and is organized as Figure~\ref{fig:AP_VSF_threshold}.

The analysis on both \vide (watershed) and threshold voids shows that void catalogs (such as the ones obtained as described in Section~\ref{subsec:vsf_cat}) are substantially unaffected by the fiducial cosmology assumed for the analysis. Interestingly, this is true for voids over all scales, including small voids, usually more strongly impacted by numerical resolution.

\bibliography{biblio_list}{}
\bibliographystyle{aasjournal}

\end{document}